\documentclass[aps,prd,preprint,groupaddress,amssymb,amsmath,nofootinbib]{revtex4}

\usepackage{color}
\usepackage{graphicx}
\usepackage{float}
\usepackage[usenames,dvipsnames]{xcolor}

\begin{document}

\preprint{}

\title{Mass scale of vectorlike matter and superpartners from IR fixed point predictions of gauge and top Yukawa couplings}

\author{Radovan Derm\' \i\v sek}

\email[]{dermisek@indiana.edu}

\author{Navin McGinnis}

\email[]{nmmcginn@umail.iu.edu}

\affiliation{Physics Department, Indiana University, Bloomington, IN 47405, USA}


\date{December 10, 2017}

\begin{abstract}
We use the IR fixed point predictions for gauge couplings and the top Yukawa coupling in the MSSM extended with vectorlike families  to infer the scale of vectorlike matter and superpartners. We quote results for several extensions of the MSSM and  present results in detail for the MSSM extended with one complete vectorlike family. We find that for a unified gauge coupling $\alpha_G > 0.3$  vectorlike matter or superpartners   are expected within 1.7 TeV (2.5 TeV) based on all three gauge couplings being simultaneously within 1.5\% (5\%) from observed values. This range  extends to about 4 TeV for $\alpha_G > 0.2$. We also find that in the scenario with two additional large Yukawa couplings of vectorlike quarks the  IR fixed point value of the top Yukawa coupling independently points to a multi-TeV range for vectorlike matter and superpartners. Assuming a universal value for all large Yukawa couplings at the GUT scale, the measured top quark mass can be obtained from the IR fixed point  for $\tan \beta \simeq 4$. The range expands to any $\tan \beta > 3$ for significant departures from the universality assumption. Considering that the Higgs boson mass also points to a multi-TeV range for superpartners in the MSSM, adding a complete vectorlike family at the same scale provides a compelling  scenario where the values of gauge couplings and the top quark mass are understood as a consequence of the particle content of the model.

\end{abstract}


\maketitle






\section{Introduction}

The gauge structure of the standard model (SM), its matter content and values of parameters might contain important clues 
for new physics. In extensions of  the SM  some of its attributes might not be just  possible choices but rather unique and some of the couplings might be related to others. 
The well known examples are supersymmetric (SUSY) grand unified theories (GUTs) that provide an understanding of many aspects of the SM 
and also lead to a successful prediction for one of the gauge couplings from a unified gauge coupling at a high scale in addition to keeping the hierarchy between the GUT scale and the electroweak (EW) scale stable~\cite{PDG_GUTs}.

Another interesting possibility is that the  values of gauge couplings are an inevitable consequence of the particle content of the theory depending very little on their boundary conditions at a high scale. This occurs in models with asymptotically divergent couplings.\footnote{Note that any model with sufficient particle content has asymptotically divergent couplings, for example, $\alpha_{EM}$ is asymptotically divergent in the SM, all three couplings are asymptotically divergent in the SM extended with 3 complete vectorlike families  and also in the MSSM extended with one complete vectorlike family.} Starting with large couplings at a high scale, in the renormalization group (RG) evolution to lower energies, couplings are driven to fixed ratios depending only on the particle content of the theory. For example, the measured value of $\sin^2 \theta_W$ was used to guess the number of families (8 to 10 chiral families) in the SM~\cite{Maiani:1977cg, Cabibbo:1982hy} before the number of chiral families and  values of gauge couplings were tightly constrained. Very good agreement between the measured value of $\sin^2 \theta_W$ and the infrared (IR) fixed point prediction of the MSSM extended with one complete vectorlike family (VF)  was noticed in Ref.~\cite{Moroi:1993}, see also recent Ref.~\cite{Carone:2017ubr}.
Similarly, if the additional particle content appears above the EW  scale, the discrepancies between the values of gauge couplings predicted from closeness to the IR fixed point and corresponding observed values  can be used to infer the mass scale of new physics, as was done for example in the SM extended by 3 complete VFs~\cite{Dermisek:2012as, Dermisek:2012ke}.

In this paper, we explore the robustness of predictions for gauge couplings in the MSSM extended with vectorlike families and use it to infer the scale of vectorlike matter and superpartners (and  the GUT scale) from the simultaneous fit to measured values of gauge couplings assuming a unified gauge coupling at the GUT scale. We quote results for several extensions of the MSSM and  present results in detail for the MSSM extended with one complete vectorlike family (MSSM+1VF). We consider scenarios with a common mass scale for vectorlike matter (or superpartners) at low energies and also scenarios where vectorlike masses (or superpartners) originate from a universal mass parameter at the GUT scale. To see the effect of different assumptions for vectorlike masses and superpartners we use 3-loop RG equations for gauge couplings that we customize to reflect 2-loop thresholds corresponding to individual particles in a given model. 

In addition,  we investigate whether the top quark mass or its Yukawa coupling can also be understood  from the IR fixed point behavior in these models. The top Yukawa coupling in the SM is not far from the stable IR fixed point of the RG equation determined by low energy values of gauge couplings~\cite{Pendleton:1981}. However, in the SM or in the MSSM (if the top quark mass is below the IR fixed point which depends on $\tan\beta$), the IR fixed point behavior is not very effective; the top Yukawa coupling approaches the fixed point very slowly. On the other hand, in models with asymptotically divergent couplings, the top quark Yukawa coupling approaches the IR fixed point very fast as a result of large gauge couplings  over  the whole energy interval,  no matter if the GUT scale boundary condition is far above or far below the fixed point~\cite{Dermisek:2012as}. Another difference from the SM or the MSSM is that vectorlike quarks can also have  large Yukawa couplings to the up-type Higgs doublet 
that affect the RG flow of the top Yukawa coupling and thus the IR fixed point prediction. We consider scenarios with  no additional Yukawa couplings, one and two additional large Yukawa couplings. We assume a universal boundary condition for all Yukawas but also discuss the variation of predictions when departing from the universality assumption.

Among our main results is the finding that for any unified gauge coupling, $\alpha_G$, larger than 0.3  vectorlike matter or superpartners   are expected within 1.7 TeV (2.5 TeV) based on all three gauge couplings being simultaneously predicted within 1.5\% (5\%) from observed values. This range  extends to about 4 TeV for $\alpha_G > 0.2$. Increasing the masses of superpartners pushes the preferred scale of vectorlike quarks and leptons down and vice versa.  More precise predictions can be made assuming a specific SUSY breaking scenario, specific origin and pattern of vectorlike masses, and specific GUT scale model with calculable GUT scale threshold corrections to gauge coupling. 
In addition, we find that in the scenario with two additional large Yukawa couplings of vectorlike quarks the  IR fixed point value of the top Yukawa coupling independently points to a multi-TeV range for vectorlike family and superpartners.  In this scenario, the measured top quark mass can be obtained from the IR fixed point value of the Yukawa coupling for $\tan \beta \simeq 4$ assuming a universal value of all large Yukawa couplings at the GUT scale and the range expands to any $\tan \beta > 3$ for significant departures from the universality assumption. Considering that the Higgs boson mass also points to a multi-TeV range for superpartners in the MSSM, adding a complete vectorlike family at the same scale offers a compelling  scenario where the values of gauge couplings and the top quark mass are understood as a consequence of the particle content of the model.\footnote{The motivation for the scale of superpartners and vectorlike matter is based completely on the measured values of gauge couplings and the top quark mass and does not take into account any  biases related to naturalness of EW symmetry breaking. Not assuming any specific SUSY breaking/mediation model, the scenarios we consider are sufficiently complex that none of the model parameters need to be selected precisely in order to obtain the required hierarchy between the EW scale and masses of superpartners~\cite{Dermisek:2016zvl, Dermisek:2017xmd}.}

From the model building point of view, adding vectorlike families is among the simplest  ways to extend  the SM or the MSSM. 
Consequently, there are many studies exploring various features of vectorlike families: examples include studies of their effects on gauge coupling unification and signatures~\cite{Babu:1996zv, Kolda:1996ea, Ghilencea:1997yr, AmelinoCamelia:1998tm, BasteroGil:1999dx}, and electroweak symmetry breaking and the Higgs mass~\cite{Babu:2008ge, Martin:2009bg, Dermisek:2016tzw}.
In addition, vectorlike fermions, not necessarily coming in complete GUT multiplets or accompanied by SUSY, are often introduced on purely phenomenological grounds to explain various anomalies. Examples include discrepancies in precision Z-pole observables~\cite{Choudhury:2001hs, Dermisek:2011xu, Dermisek:2012qx, Batell:2012ca}, and  the muon g-2 anomaly~\cite{Kannike:2011ng, Dermisek:2013gta}.

This paper is organized as follows. In Sec.~\ref{sec:RGE}, we study the IR fixed point predictions for gauge couplings and consequences for the spectrum of vectorlike matter and superpartners. In Sec.~\ref{sec:top}, we study the IR fixed point prediction for the top Yukawa coupling. We summarize and discuss results in Sec.~\ref{sec:conclusions}. The 3-loop RG equations for gauge couplings that include 2-loop threshold effects from superpartners and vectorlike matter together with two loop equations for Yukawa couplings and vertorlike masses are presented in the Appendix.

\section{Gauge couplings}
\label{sec:RGE}

The renormalization group (RG) evolution for the gauge couplings of $SU(3)\times SU(2)_L\times U(1)_Y$ is determined by the first-order differential equations,
\begin{equation}
\frac{d\alpha_i}{dt}=\beta(\alpha_i),
\end{equation}
where $\alpha_i = g_i^2/4\pi$ and $t=\ln Q/Q_0$ with $Q$  representing the energy scale at which gauge couplings are evaluated.  At one-loop level, the $\beta$ functions are simple, 
\begin{equation}
\beta(\alpha_i)=\frac{\alpha_i^2}{2\pi}b_i,
\end{equation}
where the coefficients $b_i$ depend on the particle content of the theory. We will consider extensions of the MSSM with vectorlike matter in 5 and 10 dimensional representations  of $SU(5)$, or 16 of $SO(10)$. We will use $n_5$ and $n_{10}$ to count the number of \textit{pairs} of additional multiplets, i.e. $(5\oplus\bar{5})$ and $(10\oplus\bar{10})$. For complete pairs of vectorlike families (VF), when $n_5=n_{10}$, we define $n_{16}\equiv n_5 = n_{10}$. In this convention, the one-loop $\beta$-function coefficients are
\begin{equation}
b_i = (33/5, 1, -3) + n_5(1,1,1) + 3n_{10}(1,1,1).
\label{eq:b}
\end{equation}
The MSSM beta functions are recovered for $n_5 = n_{10}=0$ and for our main example, the MSSM extended by 1 complete vectorlike family, $n_{16} = 1$, we have  $b_i = (53/5,5,1)$. Note that at one loop the beta functions for $n_{16} = 1$ and $n_{5} = 4$ are identical  and these choices represent the minimal matter content for which all three gauge couplings are asymptotically divergent.

\begin{figure}[t]
\centering
\includegraphics[width=3.in ]{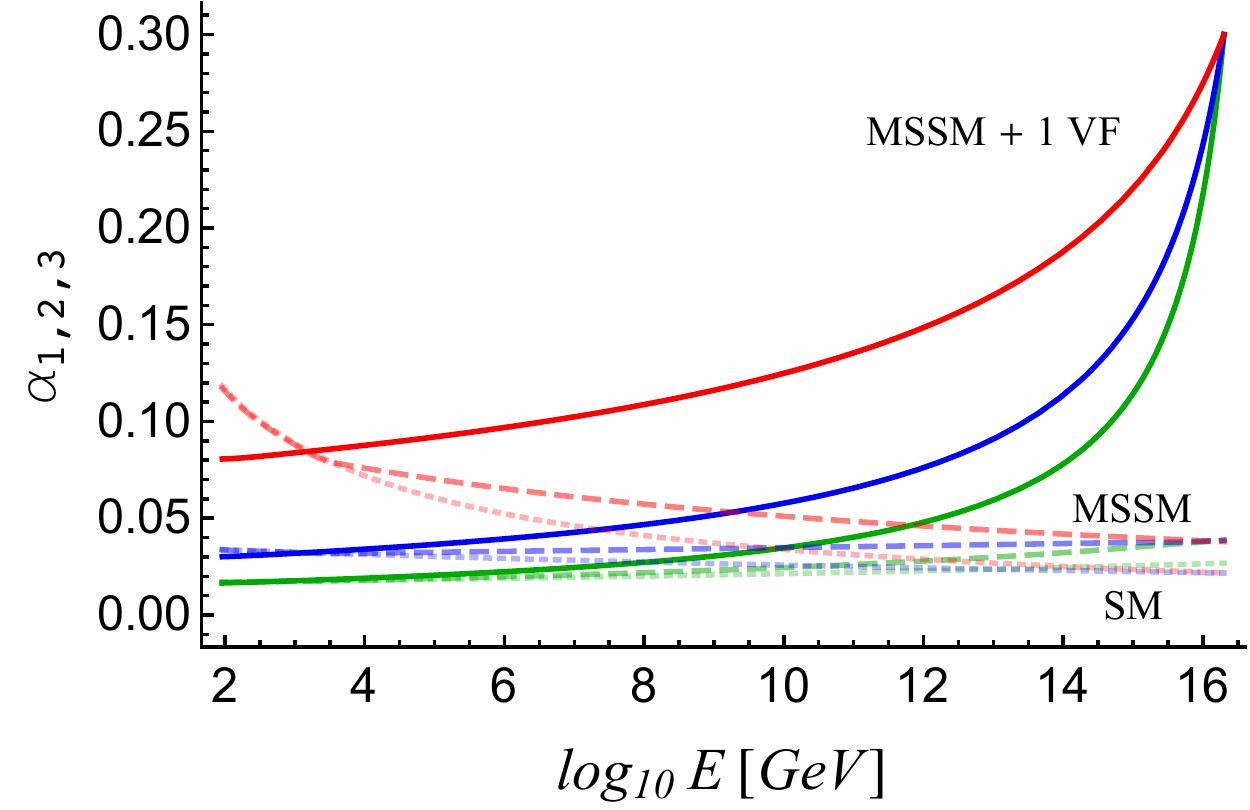}
\caption{Renormalization group evolution (3-loop) of the gauge couplings, $\alpha_3$ (top), $\alpha_2$ (middle) and $\alpha_1$ (bottom), in the SM (dotted lines), the  MSSM (dashed lines) and in the extension of the MSSM with one complete vectorlike family, $n_{16} = 1$ (solid lines). For the MSSM, we assume all superpartners at $M_{SUSY} = 3$ TeV. For the MSSM+1VF, we set $\alpha_G=0.3$ at $M_G = 2\times 10^{16}$ GeV and the full particle content is assumed all the way to the EW scale.}
\label{fig:VF_run}
\end{figure}

The evolution of gauge couplings in the SM, the MSSM  and an example of the evolution  in the MSSM+1VF are shown in Fig.~\ref{fig:VF_run}. For the SM and MSSM evolutions we have used the
central values of $\alpha_{EM}^{-1} (M_Z) = 127.916$, $\sin^2 \theta_W = 0.2313$, and $\alpha_{3} (M_Z) = 0.1184$~\cite{Patrignani:2016xqp}. The values of $\alpha_{1,2}$ are related to $\alpha_{EM}$ and $\sin^2\theta_W$ by
 \begin{gather}
 \sin^2\theta_W = \frac{\alpha^\prime}{\alpha_2 + \alpha^\prime},\nonumber\\
 \alpha_{EM} = \alpha_2\sin^2\theta_W,
 \label{eq:a_EM}
 \end{gather}
where we assume the $SU(5)$ normalization of the hypercharge, $\alpha' \equiv  (3/5)  \alpha_1$. We fix the top quark mass to 173.1 GeV, and, for the moment, we assume $\tan\beta =10$  and neglect all other Yukawa couplings. In addition, for the evolution in the MSSM, all superpartner masses are set to $M_{SUSY} =3$ TeV at the $M_{SUSY}$ scale with $A$-terms set to $-M_{SUSY}$ which is consistent with obtaining the correct mass of the Higgs boson. We use 3-loop RG equations and all particles with masses above $M_Z$ start contributing at their mass scale (see the Appendix). The RG evolution shows the well known fact that the gauge couplings approximately unify
at $M_G \simeq 2\times 10^{16}$ GeV. The example of the RG evolution of gauge couplings in the MSSM+1VF starts with a unified gauge coupling $\alpha_G = 0.3$ at the same $M_G$. The full particle content is assumed all the way to the EW scale.

The similarities and differences of the evolution of gauge couplings in these models can be qualitatively understood  from the solution of the one-loop RG equations,
\begin{equation}
\alpha_i^{-1}(M_Z)=\frac{b_i}{2\pi}\ln\frac{M_G}{M_Z} + \alpha^{-1}(M_G).
\label{eq:a}
\end{equation}
It is well known that adding complete SU(5) multiplets at a common scale does not change the scale of unification, $M_G$, at one-loop, since all three beta function coefficients increase by the same amount, see Eq.~(\ref{eq:b}). However, the unified coupling $\alpha_G$  increases with additional matter content. 
With increasing the number of vectorlike families, the couplings can become non-perturbative and   reach the Landau pole before they meet. Further increase of the number of families  lowers the energy scale at which the Landau pole occurs. 

This behavior of gauge couplings allows us to consider models with a large (but still perturbative) unified gauge coupling at a high scale, higher than the scale at which the Landau pole would occur if the VFs were at the EW scale, and use the measured values of gauge couplings to determine the mass scale of VFs. This approach was used for standard model extensions with vectorlike families~\cite{Dermisek:2012as, Dermisek:2012ke}. In the example given in Fig.~\ref{fig:VF_run}, we see that  the crossing of the evolution of gauge couplings in the MSSM+1VF and the MSSM or the SM indicates the scale of  the vectorlike family, $\gtrsim 1$ TeV, required to reproduce the measured values of all three gauge couplings.\footnote{Since the crossings of individual gauge couplings do not point to exactly the same scale, some GUT scale threshold corrections or some splitting of vectorlike masses (leading to threshold corrections near the EW scale) is required in order to reproduce the measured values of the gauge couplings precisely. We will return to this later.}

One of the most attractive features  of these models is that, in the RG evolution to lower energies, the gauge couplings run to the (trivial) IR fixed  point.  Thus, at lower energies, the values of the gauge couplings are determined only by the particle content of the theory and how far from the GUT scale we measure them. The first term in Eq.~(\ref{eq:a}) dominates and  the exact value of $\alpha_G$, or even whether the gauge couplings unify or not, becomes unimportant. Because of that, instead of one prediction of the conventional unification, we have two predictions for ratios of gauge couplings. That the ratios of gauge couplings are approximately constant  far away from the GUT scale can also  be seen in Fig.~\ref{fig:VF_run}. Higher loop effects do not alter a  mild energy dependence of the ratios. Finally, at the scale of VF, extra matter fields are integrated out and  below this scale the  gauge couplings run according to the usual RG equations of the MSSM or the SM. 
In a way, the two parameters of the conventional unification, $M_G$ and $\alpha_G$, are replaced by $M_G$ and $M_{VF} $.

\subsection{IR fixed point predictions for gauge couplings}

Let us neglect masses of VF for the moment and focus on the IR fixed point predictions for the ratios of gauge couplings. 
From Eq.~(\ref{eq:a}), it can be seen that for sufficiently large (but still perturbative) $\alpha_G$ the first term will be the dominating factor far away from the GUT scale and the ratios between couplings at the EW scale (or any other scale far from the GUT scale) can be understood in terms of their beta function coefficients,
\begin{equation}
\frac{\alpha_j(M_Z)}{\alpha_i(M_Z)}\simeq \frac{b_i}{b_j}.
\label{eq:ratio}
\end{equation}
Thus, these EW scale predictions are independent of any GUT scale parameter.

\begin{figure}[t]
\centering
\includegraphics[width = 3.in]{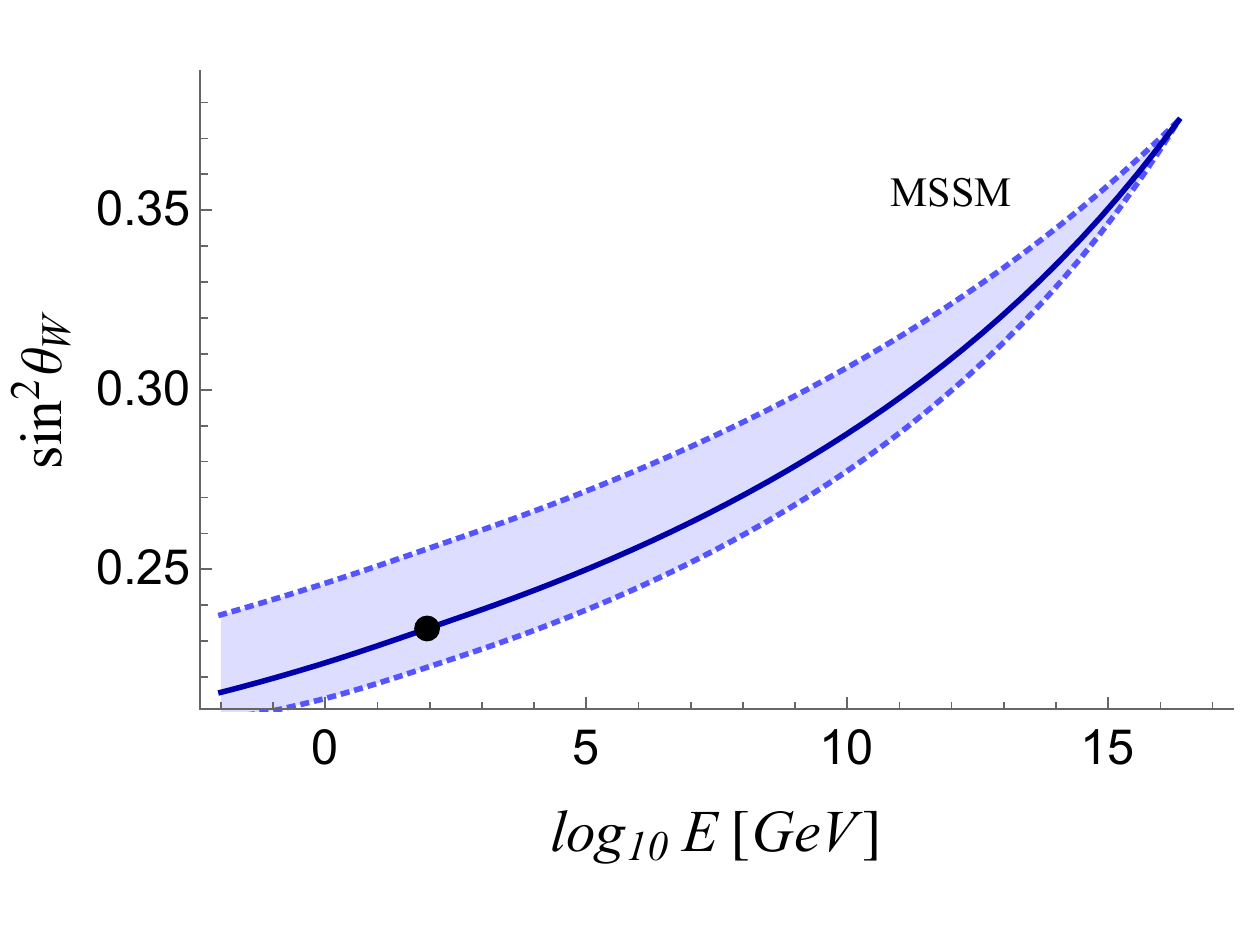}
\includegraphics[width = 3.in]{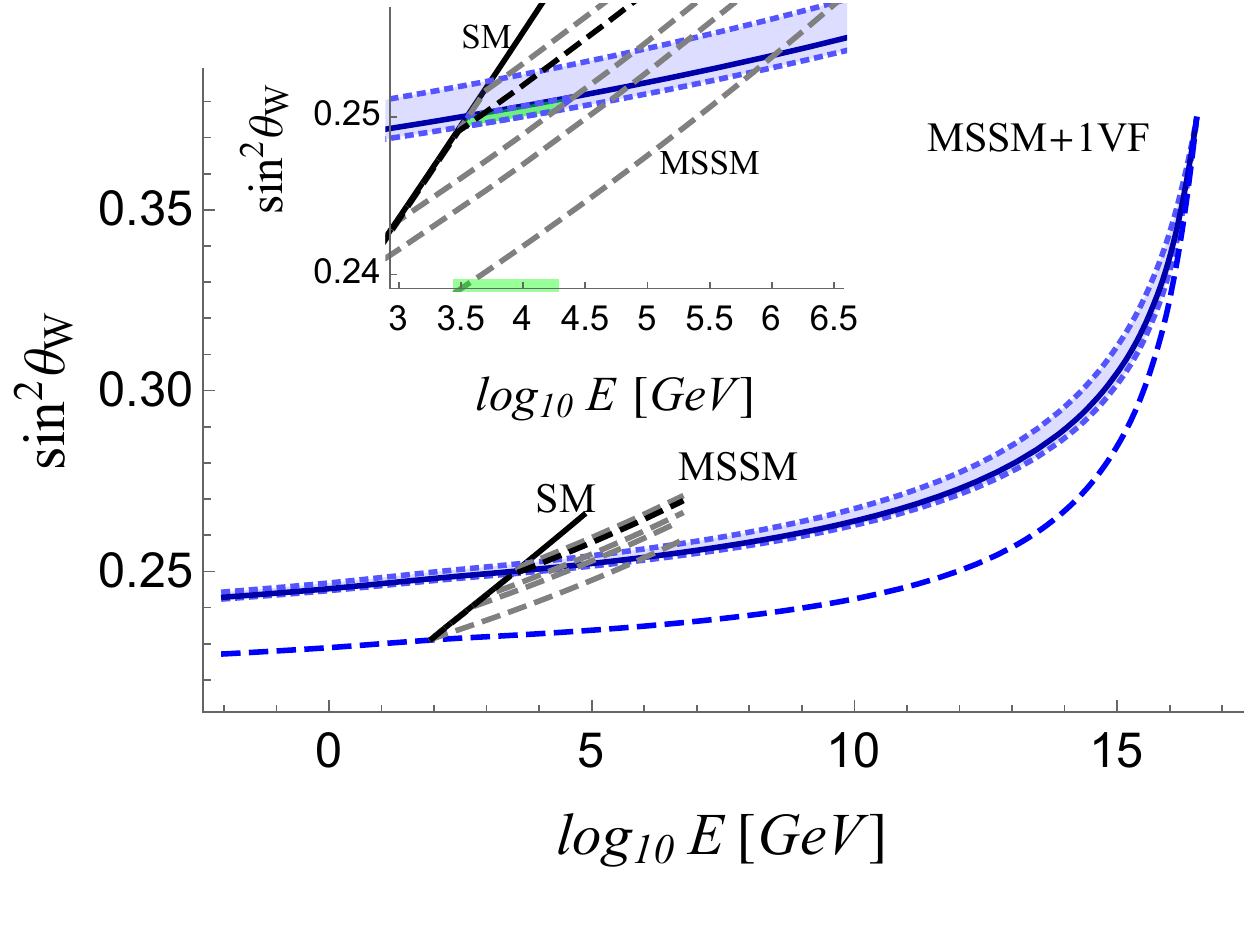}
\hspace{1cm}(a)\hspace{7cm} (b)
\caption{RG evolution of $\sin^2\theta_W$ in the MSSM (a) and MSSM+1VF (b) neglecting threshold effects from superpartners and VF. In (a), the dot shows the measured  value of 
$\sin^2\theta_W(M_Z)$, the solid line  represents its evolution according to 3-loop MSSM RG equations and dotted lines (and shaded region) illustrate the changes in the prediction resulting from varying $\alpha_G$ in the $\pm 30\%$ range around the MSSM value. In (b), the evolution of $\sin^2\theta_W$ is shown for $\alpha_G = 0.3$ at one-loop (dashed line) and 3-loop (solid line). The variation of the 3-loop prediction resulting from varying  $\alpha_G$ in the $\pm 30\%$ range is illustrated by dotted lines and shaded region.  At low energies we also show the RG evolution of $\sin^2\theta_W$ in the SM (solid black) and in the MSSM (dashed) with all superpartners at $M_Z$ (bottom dashed line), 500 GeV, 1 TeV, 3 TeV (black dashed line) and 5 TeV (top dashed line). The inset zooms in the region at low energies and the highlighted energy range indicates  the masses of the vectorlike family needed for reproducing the measured value of $\sin^2\theta_W(M_Z)$ for any $\alpha_G > 0.3$ and superpartners above 1 TeV.}
\label{fig:weinberg_run}
\end{figure}

For example, this relation between couplings can be used to obtain a prediction for the Weinberg angle, 
\begin{equation}
\sin^2\theta_W = \frac{\alpha^\prime}{\alpha_2+\alpha^\prime}\simeq \frac{b_2}{b^\prime + b_2},
\label{eq:sin2W}
\end{equation}
where $b^\prime = (5/3)b_1$. For the MSSM+1VF, this gives $\sin^2\theta_W \simeq 0.2205$ which is within $5\%$ of its observed value.  The virtue of this prediction can be seen in Fig.~\ref{fig:weinberg_run} where we show the RG evolution of $\sin^2\theta_W$  in the MSSM (a) and in the MSSM+1VF (b). In the MSSM, the predicted value of $\sin^2\theta_W$  crucially depends on the GUT scale and it varies significantly with changes in $\alpha_G$.  In contrast, for the MSSM+1VF we see that  $\sin^2\theta_W$ has essentially the same value in a huge range of the energy scale, away from the GUT scale, and is almost unchanged for comparable variations in $\alpha_G$. Higher loop effects slightly increase the predicted value, however the insensitivity to both the GUT scale and $\alpha_G$ remains. We do not show 2-loop results since there is no visible difference between 2-loop and 3-loop results. The one-loop and 3-loop predictions for $\sin^2\theta_W$ in several extensions of the MSSM with vectorlike families are summarized in Table~\ref{tb:s2w_IR}.

\begin{table}[t]
\centering
\begin{tabular}{|c| |c|c|c||c|c|c|}
\hline
Model&$(\sin^2\theta_W)^{1-\text{loop}}_{\text{IR}}$ &$(\sin^2\theta_W)^{3-\text{loop}}_{3\,\text{TeV}}$& $M_{VF}$(GeV) &$(\alpha_3/\alpha_{EM})^{1-\text{loop}}_{\text{IR}}$& $(\alpha_3/\alpha_{EM})^{(3-\text{loop})}_{3\,\text{TeV}}$&$M_{VF}$(GeV)\\
\hline
$n_{16}=1$&0.2205&0.2485&$4.44\times 10^{3}$ &22.66&10.59&$1.27\times 10^3$\\
\hline
$n_{16}=2$ &0.2700&0.2857&$5.15\times 10^{9}$&6.66&5.64&$2.97\times 10^9$\\
\hline
$n_5=4$&0.2205&0.2429&$257$&22.66&10.71&$1.03\times 10^3$\\
\hline
$n_5=5$ &0.2368&0.2550&$1.10\times 10^{5}$&12.66&8.37&$2.81\times 10^5$\\
\hline
$n_{10}=2$&0.2500&0.2719&$6.52\times 10^{7}$&9.33&7.00&$1.91\times 10^7$\\
\hline
$n_{10}=3$&0.2777&0.2933& $3.70\times 10^{10}$&6.00&5.23&$1.77\times 10^{10}$\\
\hline
\end{tabular}
\caption{One-loop and 3-loop predictions for $\sin^2\theta_W$  and $\alpha_3/\alpha_{EM}$ in various extensions of the  MSSM. The one-loop results represent the IR fixed point predictions, Eqs.~(\ref{eq:sin2W})  and (\ref{eq:3oEM}). The 3-loop results represent predictions at the 3 TeV scale starting from $\alpha_G = 0.3$ at $M_G = 3\times 10^{16}$ GeV. The $M_{VF}$ represents the common scale for vectorlike masses inferred from the observed values of either $\alpha_3/\alpha_{EM}$ or $\sin^2\theta_W$ assuming all superpartners at 3 TeV. Note that  in the SM we have  $\sin^2\theta_W(3\,\text{TeV})=0.2491$  and  $\alpha_3/\alpha_{EM}(3\,\text{TeV})=10.04$.}
\label{tb:s2w_IR}
\end{table}

We can gain some indication of the decoupling scale for vectorlike matter if we compare the running of this parameter with that in the SM and the MSSM at low energies.  In the inset of Fig.~\ref{fig:weinberg_run}(b), we can see that the crossing of the evolution of $\sin^2\theta_W$ in the SM and in the MSSM+1VF appears around 3 TeV.  Assuming comparable masses for superpartners and vectorlike fields, in order to obtain the correct value of $\sin^2\theta_W(M_Z)$,  all the extra matter must be decoupled near this scale. For lighter superpartners, the evolution of $\sin^2\theta_W$ in the MSSM (dashed lines) crosses the evolution of $\sin^2\theta_W$  in the MSSM+1VF at higher energies. However, for any $\alpha_G > 0.3$ and superpartners above 1 TeV the indicated common scale for vectorlike matter is below  20 TeV.
We will explore the needed scale for superpartners and vectorlike matter in more detail in the following subsection.

\begin{figure}[t]
\centering
\includegraphics[width = 3.in]{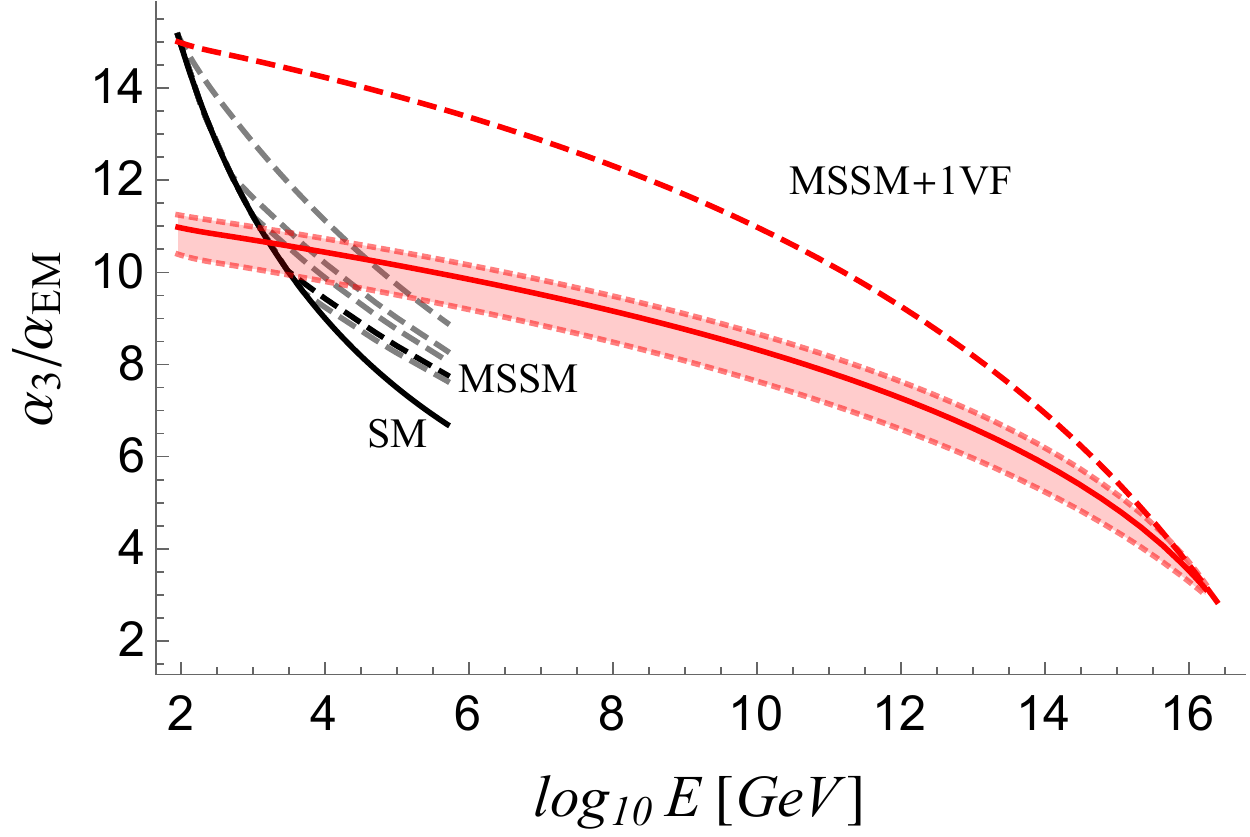}
\includegraphics[width = 3.in]{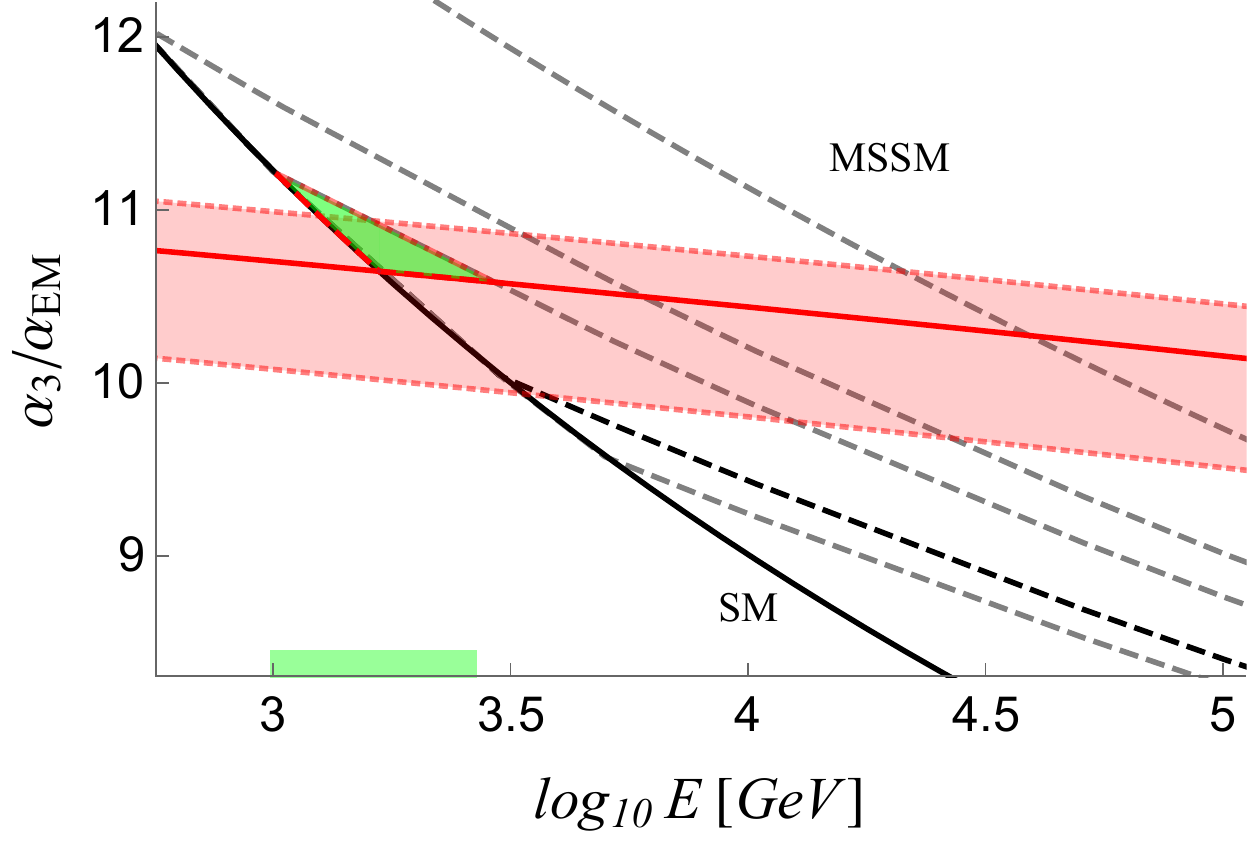}\\
\hspace{1cm}(a)\hspace{7cm} (b)
\caption{(a) RG evolution of $\alpha_3/\alpha_{EM}$ in the MSSM+1VF  neglecting threshold effects from superpartners and VF for $\alpha_G = 0.3$ at one-loop (dashed line) and 3-loop (solid line). The variation of the 3-loop prediction resulting from varying  $\alpha_G$ in the $\pm 30\%$ range is illustrated by  dotted lines and shaded region.  At low energies we also show the RG evolution of $\alpha_3/\alpha_{EM}$ in the SM (solid black) and in the MSSM (dashed) with all superpartners at $M_Z$ (top dashed line), 500 GeV, 1 TeV, 3 TeV (black dashed line) and 5 TeV (bottom dashed line). (b) Shows a zoomed in region at low energies and the highlighted energy range indicates  the masses of vectorlike family needed for reproducing the measured value of $\alpha_3/\alpha_{EM}$ for any $\alpha_G > 0.3$ and superpartners above 1 TeV. }
\label{fig:a3_aem}
\end{figure}

Another parameter free prediction of the model can be obtained for the ratio $\alpha_3/\alpha_{EM}$ by combining Eqs.~(\ref{eq:a_EM}), (\ref{eq:ratio}), and (\ref{eq:sin2W}). At the one-loop level, far below the GUT scale we have
\begin{equation}
\frac{\alpha_3}{\alpha_{EM}}= \frac{b_2}{b_3}\frac{1}{\sin^2\theta_W} = \left(\frac{b_2+b^\prime}{b_3}\right).
\label{eq:3oEM}
\end{equation}
We can obtain similar one-loop predictions as for $\sin^2\theta_W$ based purely on group theoretical factors and particle content. These, together with 3-loop predictions, are summarized in Table~\ref{tb:s2w_IR} for various extensions of the MSSM with vectorlike families. The 1-loop prediction is typically not a very good approximation, especially for $n_5 = 4$ and $n_{16}=1$ cases, since the beta-function coefficient $b_3$ is small and thus 2-loop effects are large. With increasing the numbers of families the one-loop predictions are getting closer to three-loop predictions.

The observed value, $\alpha_3/\alpha_{EM}(M_Z)=15.14$, is far from any of the predictions. However, as we can see from Fig.~\ref{fig:VF_run}, $\alpha_3$ in the SM runs fast at low energies while $\alpha_{EM}$ does not. For example, already at 3 TeV we have $\alpha_3/\alpha_{EM}(3 \, \rm{TeV})=10.04$, which is in good agreement with predictions of models with $n_{16}=1$ or $n_5 = 4$. The common scales of vectorlike families that lead to  observed values of $\alpha_3/\alpha_{EM}$ and $\sin^2\theta_W$ are also indicated in Table~\ref{tb:s2w_IR}.

The RG evolution of $\alpha_3/\alpha_{EM}$  in the MSSM+1VF is shown in Fig.~\ref{fig:a3_aem}. We see that higher loop effects are indeed more important in this case and, also due to small $b_3$, the  $\alpha_3/\alpha_{EM}$ is approaching the IR fixed point prediction much slower compared to the $\sin^2\theta_W$. Nevertheless,  the insensitivity of the prediction to both the GUT scale and $\alpha_G$ far below the GUT scale is still significant. There is again  no visible difference between 2-loop and 3-loop results. From the crossing of the evolutions of  this parameter in the MSSM+1VF and in the SM  we see  that a common scale of superpartners and vectorlike fields should be around  2 TeV.
 For lighter superpartners, the evolution of $\alpha_3/\alpha_{EM}$ in the MSSM (dashed lines) crosses the evolution of $\alpha_3/\alpha_{EM}$  in the MSSM+1VF at higher energies. Thus, for any $\alpha_G > 0.3$ and superpartners anywhere above 1 TeV the indicated common scale for vectorlike matter is below  2.6 TeV. In what follows, we will explore predictions of the MSSM+1VF model in more detail.

\subsection{Scale of vectorlike matter and superpartners  in the MSSM+1VF}

\begin{figure}[t]
\centering
\begin{minipage}[b]{0.4\linewidth}
\includegraphics[width=3.in]{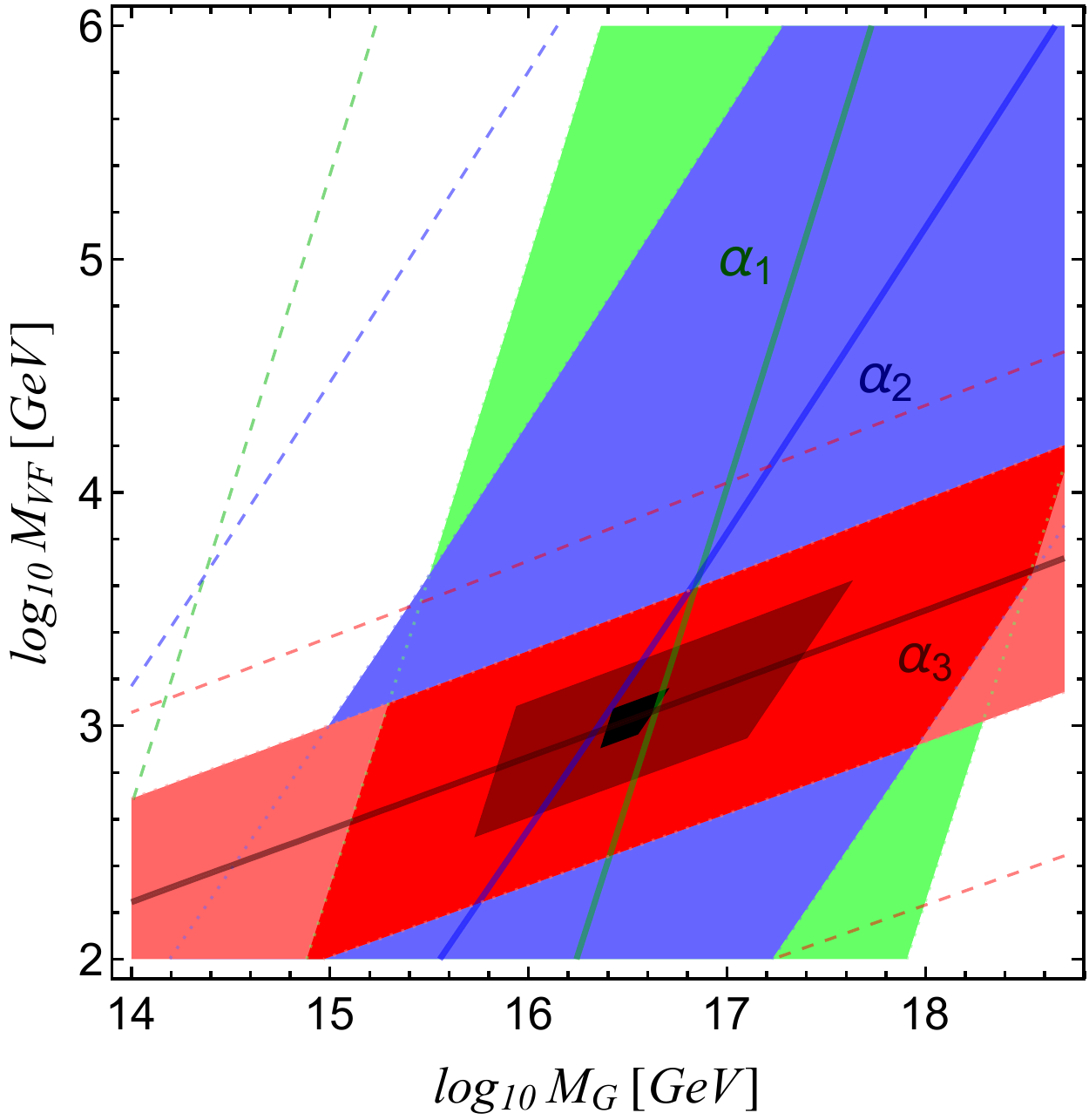}\\
\hspace{2.cm}(a) 
\end{minipage}
\begin{minipage}[b]{0.45\linewidth}
\centering
\hfill\includegraphics[width=2.in]{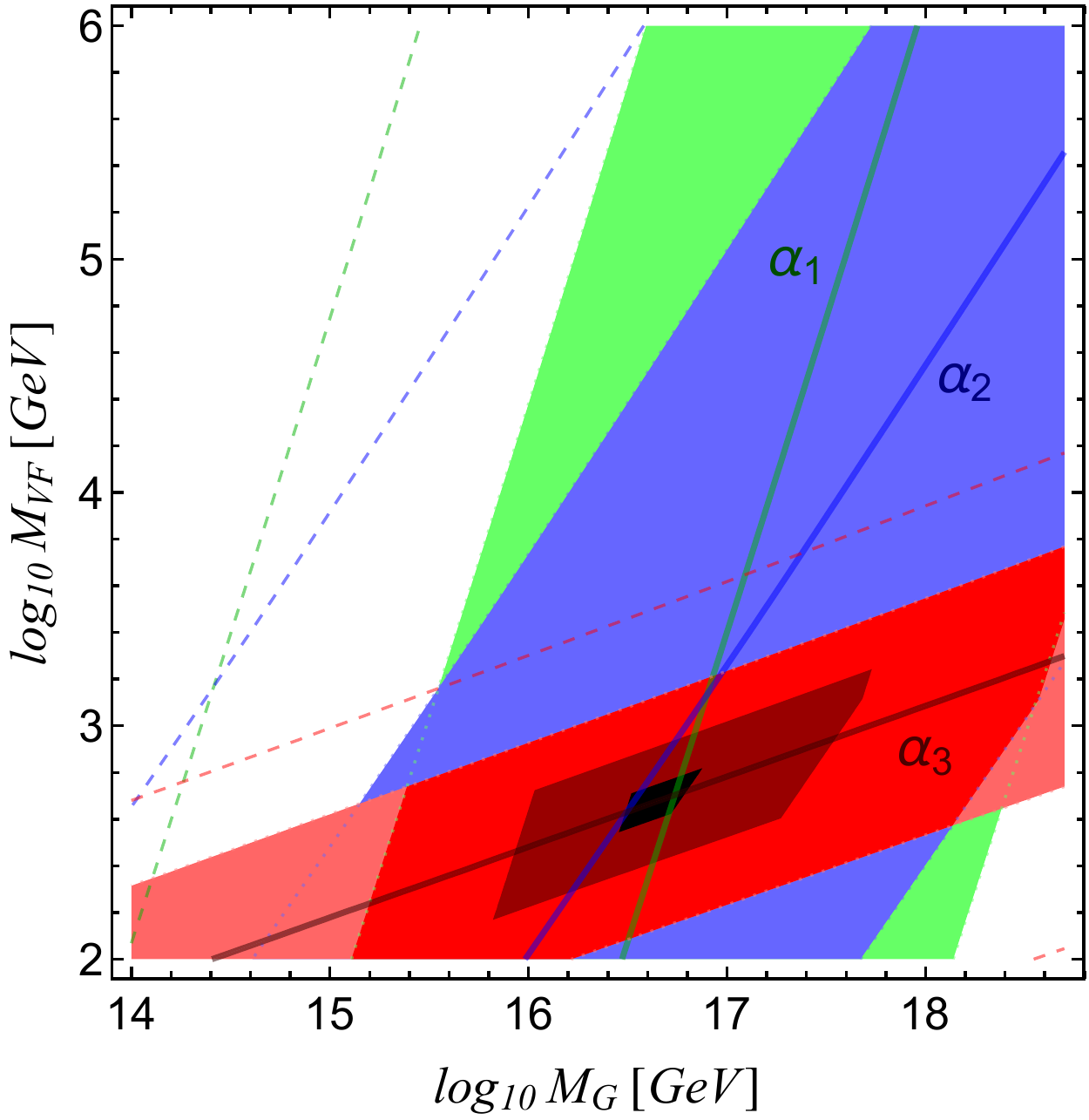}\\ \hspace{2.5cm}(b)\\
\vspace{0.1cm}
\hfill\includegraphics[width=2.in]{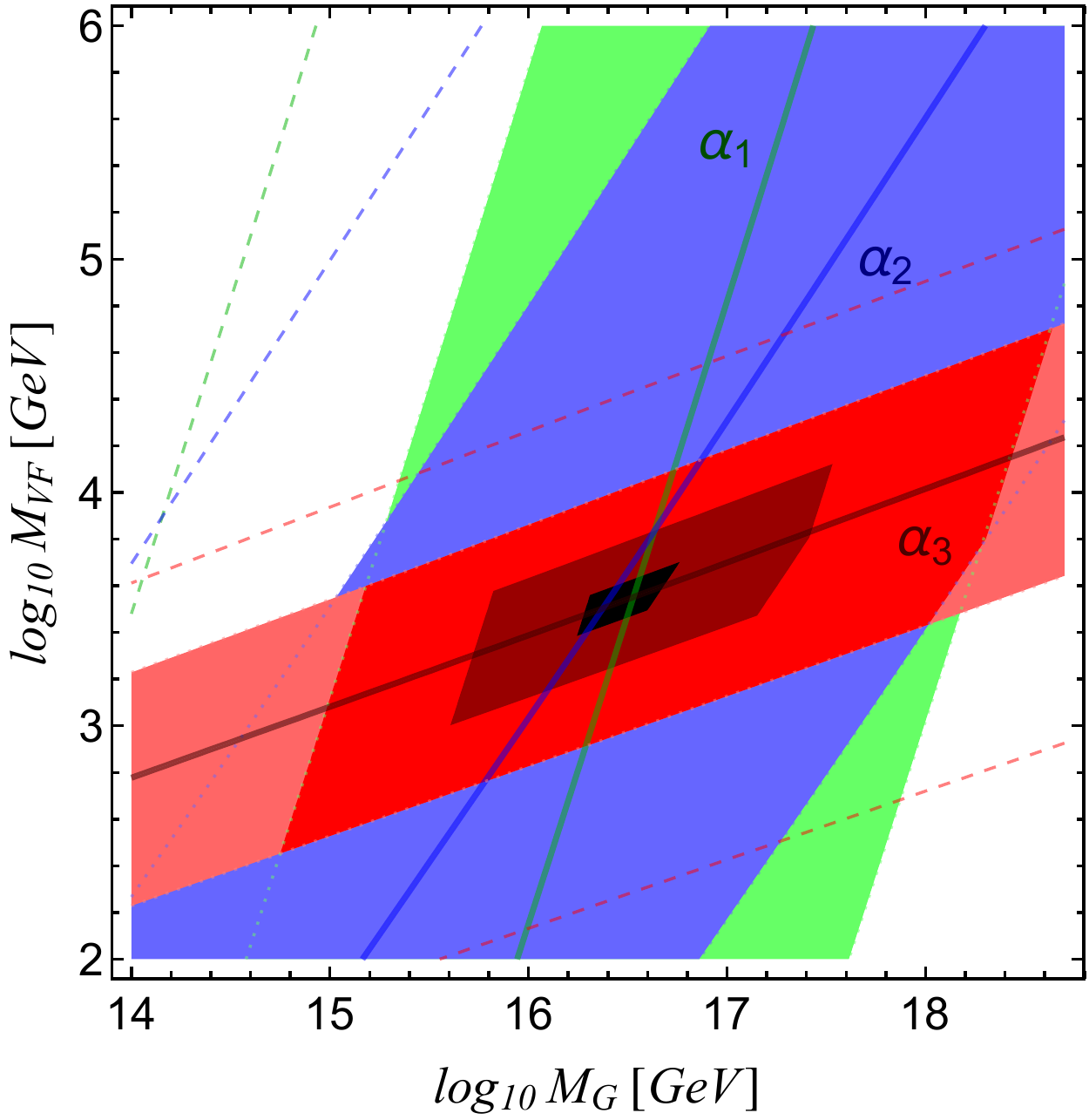}\\ \hspace{2.5cm}(c)
\end{minipage}
\caption{Contours of constant values of predicted gauge couplings at $M_Z$, $\alpha_1$ (green),  $\alpha_2$ (blue), and $\alpha_3$ (red), as functions of $M_{G}$ and the universal mass of vectorlike family, $M_{VF}(M_{VF})$,  in the MSSM+1VF for three values of $\alpha_G$:  $\alpha_G = 0.3$ (a), $\alpha_G = 0.4$ (b), and $\alpha_G = 0.2$ (c). All superpartners are integrated out at a common scale 3 TeV. Solid lines represent the central experimental values of three gauge couplings, the shaded regions represent $\pm 10\%$ ranges, and the dashed lines represent $\pm20\%$ ranges. In the overlapping (bright red) region,  all three gauge couplings are simultaneously predicted within $10\%$ from the measured values. In the smaller dark red and black regions, all three couplings are simultaneously predicted within $\pm5\%$ and $\pm 1.5\%$ respectively.}
\label{fig:MVF_SUSY_3TeV}
\end{figure}

Let us start with the assumption of a common mass  scale for vectorlike matter, $M_{VF}$ (at the $M_{VF}$ scale). This parameter together with the GUT scale are the most important determining factors for gauge couplings at the EW scale. Predictions for gauge couplings at the EW scale as functions of  $M_G$ and $M_{VF}$, using 3-loop RG equations, are shown in Fig.~\ref{fig:MVF_SUSY_3TeV} for fixed values of the unified gauge coupling: $\alpha_G = 0.3$ (a), $\alpha_G = 0.4$ (b) and $\alpha_G = 0.2$ (c). In this figure, a common scale for all superpartner masses, $M_{SUSY} =3$ TeV (at the 3 TeV scale), is assumed which corresponds to the black dashed lines in Figs.~\ref{fig:weinberg_run} and \ref{fig:a3_aem} (the same scale was also assumed in Table~\ref{tb:s2w_IR}). For top soft trilinear coupling $A_t \simeq - M_{SUSY}$, the spectrum is consistent with the measured value of the Higgs  boson mass.

\begin{figure}[t]
\centering
\includegraphics[width=3.in]{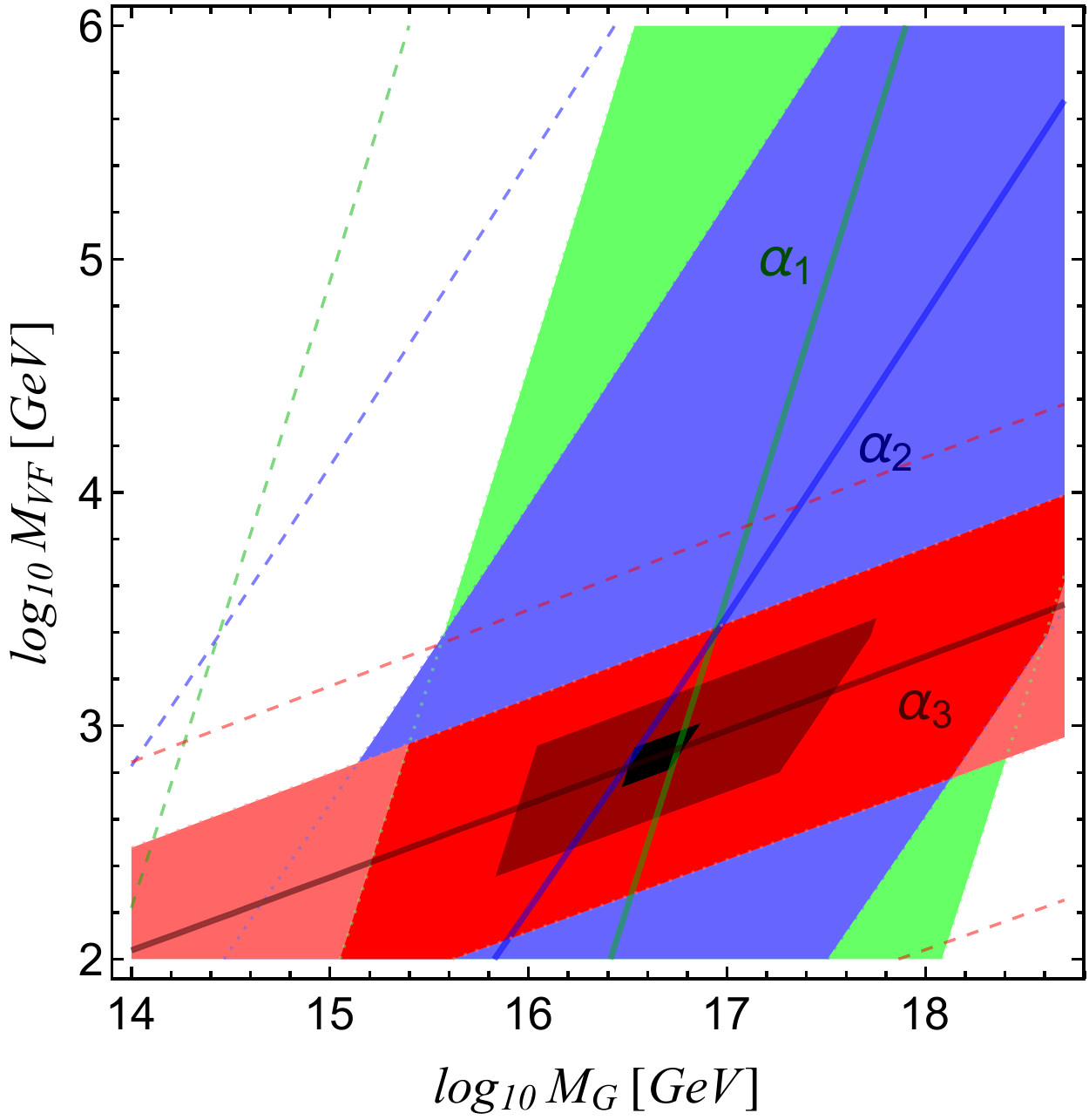} \hspace{0.5cm}
\includegraphics[width = 3.in]{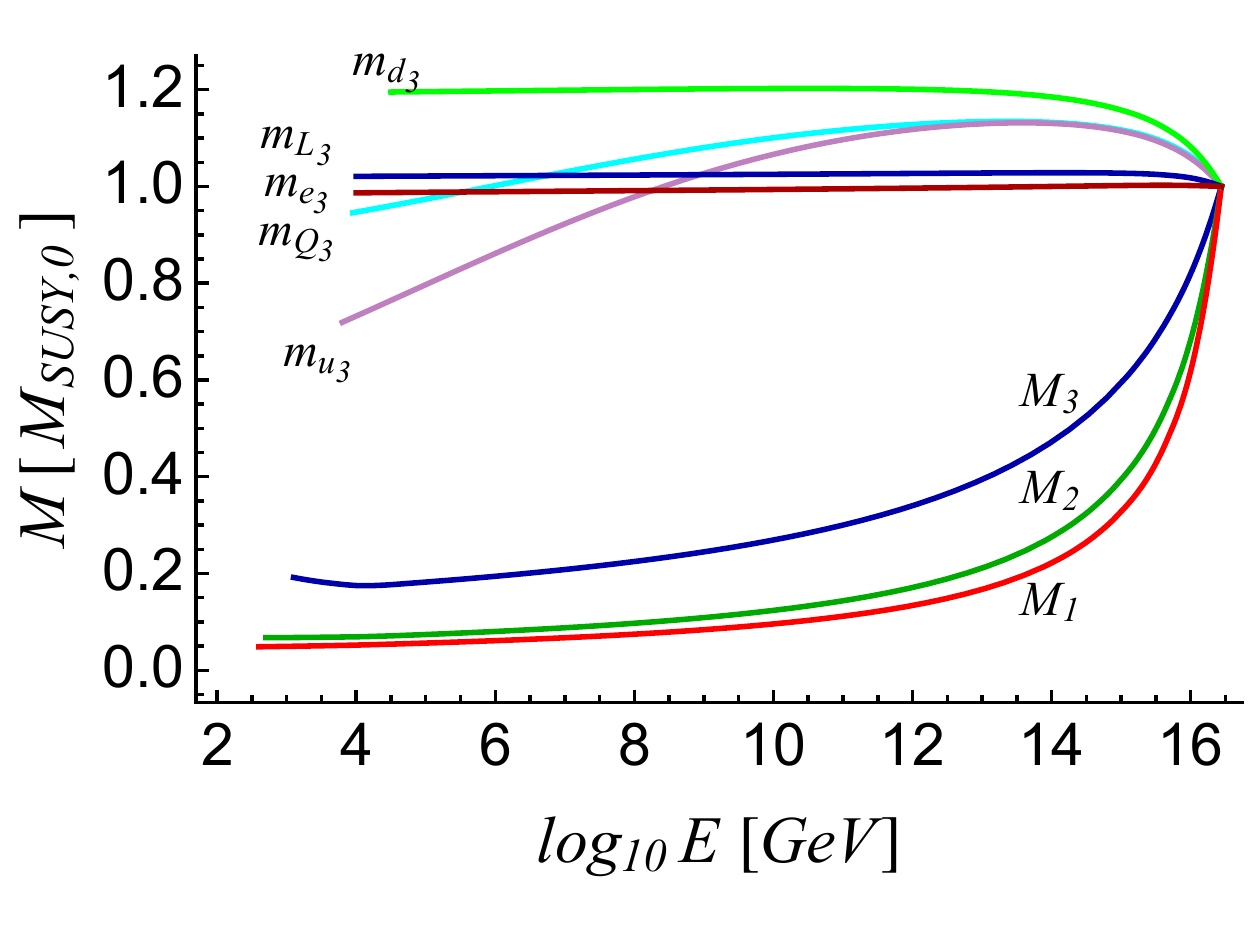}
\hspace{1cm}(a)\hspace{8cm} (b)
\caption{(a) The same as in Fig.~\ref{fig:MVF_SUSY_3TeV}(a) but with superpartners integrated out at their corresponding mass scales resulting from $M_{SUSY,0}=9$ TeV at the GUT scale for   $\alpha_G = 0.3$. (b) 2-loop RG evolution of gaugino and 3rd generation scalar masses in the MSSM+1VF for $\tan \beta = 10$ starting from  $M_{SUSY,0}\equiv M_{1/2} = m_0$ at the GUT scale with masses normalized to  $M_{SUSY,0}$. The evolution of a given parameter stops at the corresponding mass scale for $M_{SUSY,0} = 9$ TeV.}
\label{fig:MVF_SUSY_9TeV}
\end{figure}

A similar plot, but assuming universal soft SUSY breaking mass parameters at the GUT scale, $M_{SUSY,0}\equiv M_{1/2} = m_0=9$ TeV, is given in Fig.~\ref{fig:MVF_SUSY_9TeV}(a).  In this case, superpartner masses are determined from 2-loop RG evolution  and they stop contributing to the running of the gauge couplings at their corresponding scales, see the Appendix for the RG equations including two loop threshold effects.  The spectrum of superpartners obtained from $M_{SUSY,0}=9$ TeV is shown in Fig.~\ref{fig:MVF_SUSY_9TeV}(b).  It satisfies limits from direct searches ($M_{SUSY,0}=9$ TeV is motivated mainly by the limits on the gluino mass) and is  consistent with the measured value of the Higgs  boson mass.\footnote{The SUSY spectrum is very different in this model compared to the MSSM. Among  interesting features is the closeness of $M_1$ and $M_2$ at low energies. Thus, a small departure from the universality assumption at the GUT scale can lead to Wino being the lightest supersymmetric particle.}

Focusing on the black spots  in Figs.~\ref{fig:MVF_SUSY_3TeV}(a) and \ref{fig:MVF_SUSY_9TeV} we see that the 
 scale of unification giving the best prediction for all three gauge couplings is essentially unchanged, as expected. More importantly, decoupling the vectorlike content at $M_{VF}\simeq 1$ TeV, assuming universal superpartner masses at 3 TeV, or the spectrum obtained from universal GUT scale values of soft parameters,  $M_{SUSY,0}= 9$ TeV, results in all the gauge couplings within $1.5\%$ of their measured values. Increasing superpartner masses requires smaller $M_{VF}$. Furthermore, these predictions are not very sensitive to $\alpha_G $ as can be seen from Figs.~\ref{fig:MVF_SUSY_3TeV}(b) and \ref{fig:MVF_SUSY_3TeV}(c). Increasing $\alpha_G $ again requires smaller $M_{VF}$ and thus for any $\alpha_G > 0.3$ the best motivated scale of vectorlike matter is around 1 TeV. Lowering $\alpha_G $ to 0.2 increases this scale to  4 TeV. Interestingly,  the scale of superpartners suggested by the Higgs boson mass is also in a multi-TeV range.

The gauge couplings can be reproduced precisely if GUT scale threshold corrections leading to about $20\%$ splitting of individual couplings at the GUT scale  are assumed.
 This is a very similar result to the usual $3\%$ correction needed in the MSSM, since the GUT scale threshold corrections are proportional to $\alpha_G$ which in our case is about 7 times larger. Alternatively, gauge couplings can  also be reproduced precisely by splitting individual vectorlike masses within a factor of five between the lightest and heaviest.

Perhaps the most intriguing feature of this scenario is the robustness of predictions for the gauge couplings. We see that predicted values of all gauge couplings are  within $10\%$ of their measured values (bright red)  in the range of $M_G$ that spans 3 orders of magnitude and  in the range of $M_{VF}$ that spans one order of magnitude. Furthermore, even the range of $M_G$ and  $M_{VF}$ reproducing all gauge couplings within $5\%$ of their measured values (dark red) is significant. The $\alpha_3$ is the most constraining coupling because it runs fastest below the scale of vectorlike matter.

\begin{figure}[t]
\centering
\includegraphics[width = 3.in]{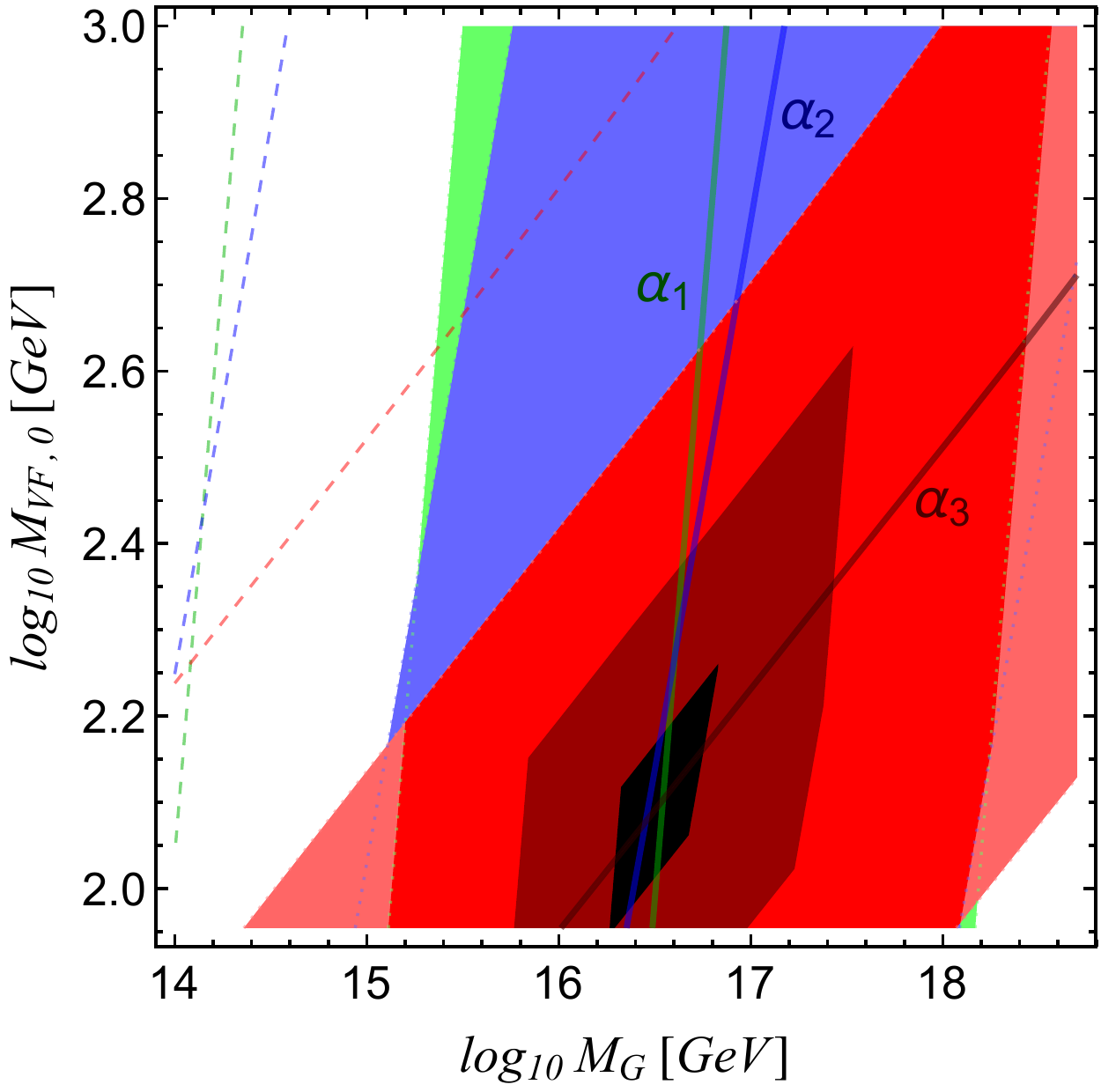} \hspace{0.5cm}
\includegraphics[width = 3.in]{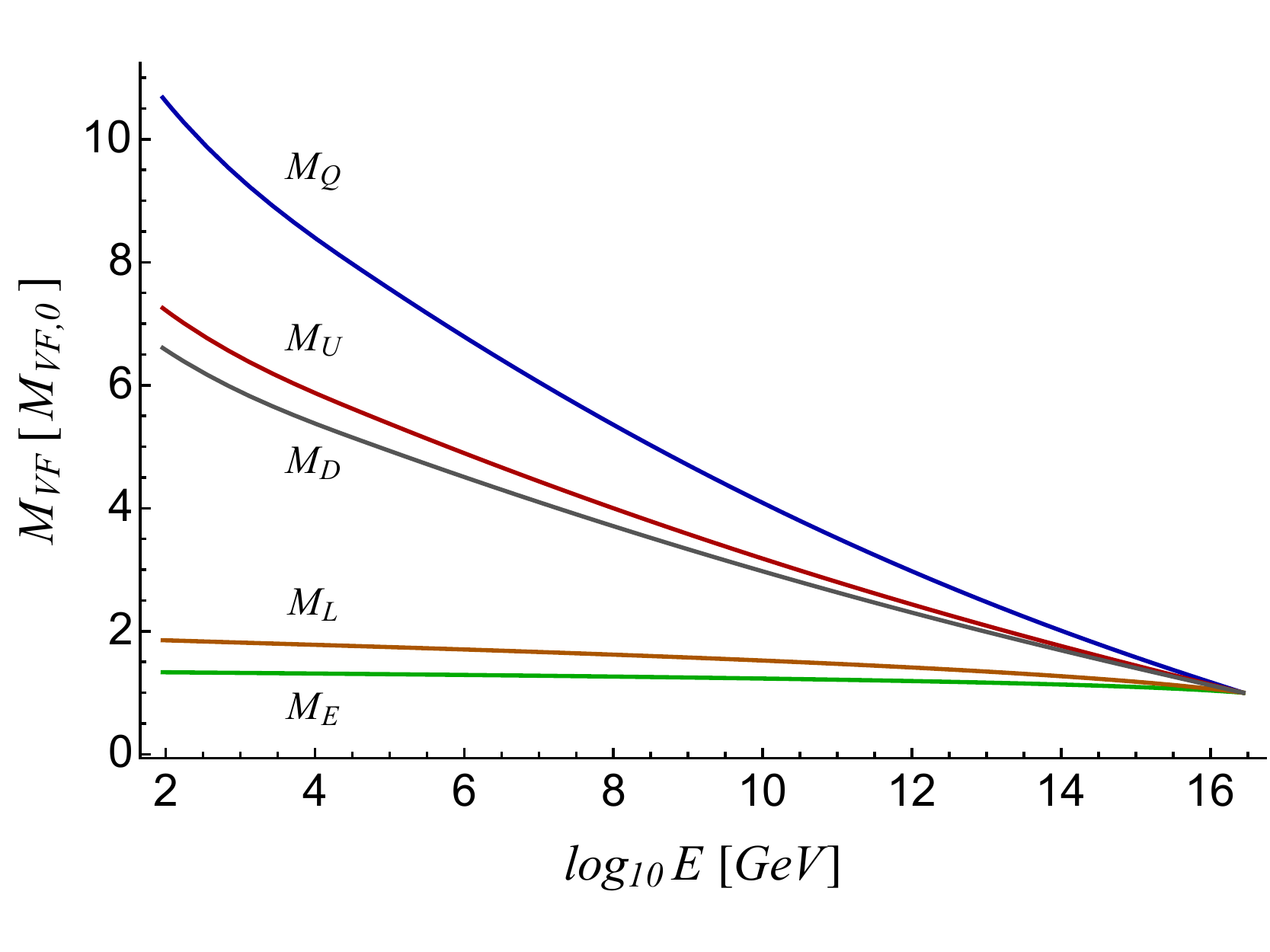}
\hspace{1cm}(a)\hspace{8cm} (b)
\caption{(a) Similar to Fig.~\ref{fig:MVF_SUSY_3TeV}(a) but with the universality condition on vectorlike masses imposed at the GUT scale, $M_{VF,0}$.  Vectorlike fields are integrated out at their corresponding mass scale.  (b) 2-loop RG evolution of vectorlike masses normalized to the GUT scale value $M_{VF,0}$.}
\label{fig:MVF0}
\end{figure}

Comparing Figs.~\ref{fig:MVF_SUSY_3TeV} and~\ref{fig:MVF_SUSY_9TeV} we see that whether superpartners are integrated at a common scale or at their corresponding masses resulting from a universal boundary condition at the GUT scale does not affect predictions for gauge couplings significantly. However, imposing the universality condition for vectorlike masses at the GUT scale instead of a low energy scale has a more dramatic effect. 
The predicted values of the gauge couplings as functions of $M_G$ and the universal vectorlike mass  at the GUT scale, $M_{VF,0}$, are shown in Fig.~\ref{fig:MVF0}(a). Other parameters are the same as in Fig.~\ref{fig:MVF_SUSY_3TeV}(a). Comparing the two plots we see that threshold effects from integrating out the vectorlike fields at their corresponding masses, resulting from the RG flow from a common mass, significantly shrinks the triangle of intersections of two individual couplings leading to predictions that agree better with observed values. Thus, significantly smaller GUT scale threshold corrections are required to precisely reproduce measured values. This  improvement  originates from almost an order of magnitude splitting between individual vectorlike masses at low energies, see Fig.~\ref{fig:MVF0}(b).

The best motivated $M_{VF,0}$ from measured values of the gauge couplings is slightly above 100 GeV which means that the vectorlike leptons are at about 200 GeV and vectorlike quarks in a TeV range. 
Vectorlike leptons with these masses are highly constrained~\cite{Dermisek:2014qca} and the vectorlike quarks at 1 TeV are near the experimental limits. However, decreasing $\alpha_G$ results in an almost identical plot with all lines moved up (as we saw in Fig.~\ref{fig:MVF_SUSY_3TeV}). For example, for $\alpha_G = 0.2$ the center of the best motivated region moves to $M_{VF,0} \simeq 1$ TeV.

\begin{figure}[t]
\centering
\includegraphics[width = 3in]{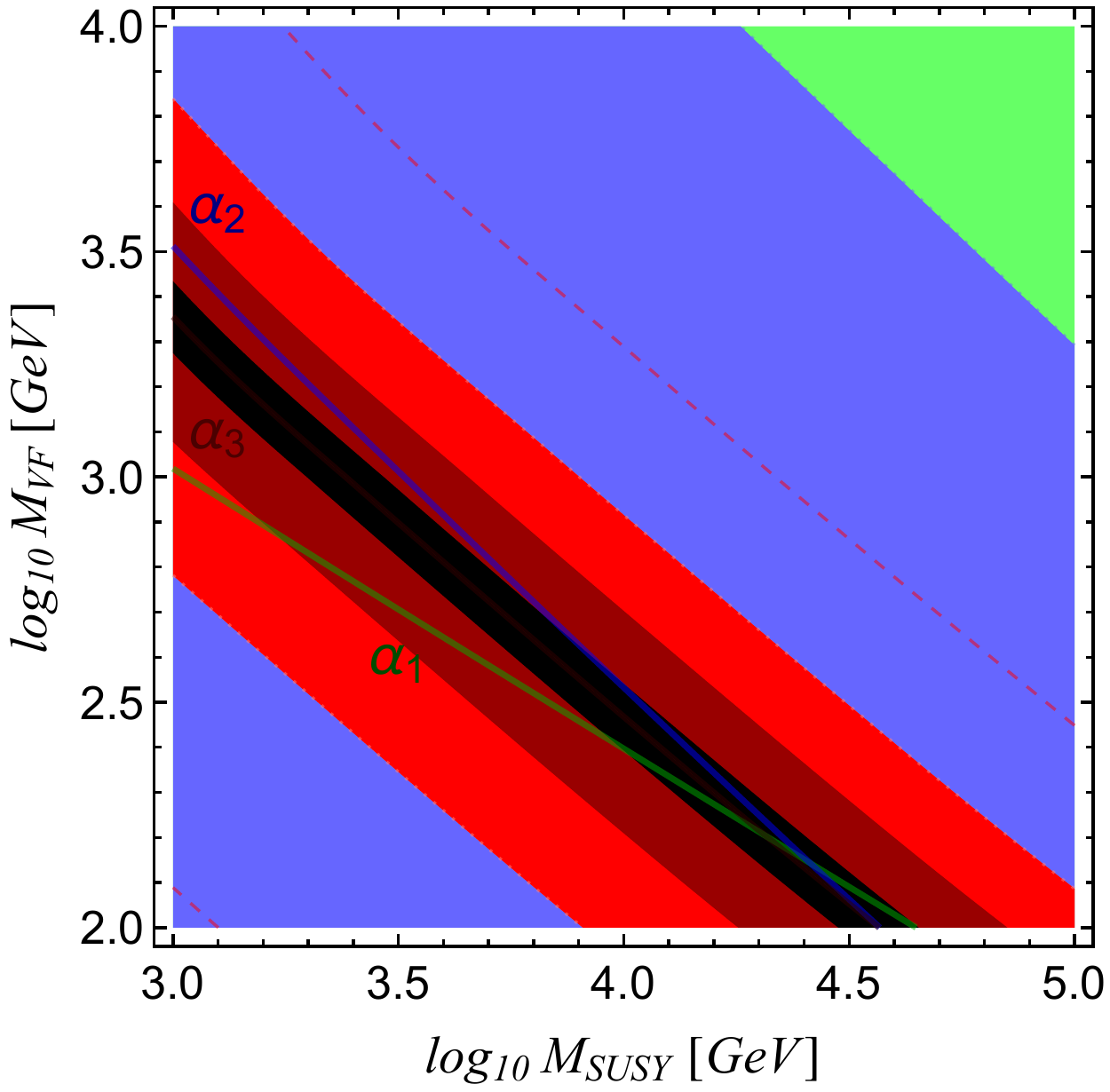}
\caption{Predicted values of gauge couplings as functions of $M_{VF}$ and $M_{SUSY}$ for $\alpha_G=0.3$ and  $M_G$ corresponding to the best fit in Fig.~\ref{fig:MVF_SUSY_3TeV}. The meaning of lines, regions and colors is the same as in Fig.~\ref{fig:MVF_SUSY_3TeV}.}
\label{fig:VF_MSUSY}
\end{figure}

Finally, we study the sensitivity of the above results with respect to $M_{SUSY}$. For $\alpha_G=0.3$ and  $M_G$ corresponding to the best fit in Fig.~\ref{fig:MVF_SUSY_3TeV}(a), we present the predicted values of gauge couplings as functions of  $M_{VF}$ and $M_{SUSY}$ (common masses at low energies) in Fig.~\ref{fig:VF_MSUSY}. We see that either superpartners or the vectorlike quarks and leptons are expected within 1.7 TeV (2.5 TeV) based on all three gauge couplings being simultaneously within 1.5\% (5\%) from their observed values. Similar  conclusions would be reached if we considered common masses of vectorlike matter or superpartners at the GUT scale. The only remaining parameter, $\alpha_G$, moves the whole plot slightly along the diagonal, and the effect of varying  $\alpha_G$ can be inferred from Figs.~\ref{fig:MVF_SUSY_3TeV}(b) and~\ref{fig:MVF_SUSY_3TeV}(c). Lowering  $\alpha_G$ to 0.2, vectorlike matter or  superpartners are expected within about 4 TeV.

For the MSSM+1VF and several other extensions of the MSSM with asymptotically divergent couplings, we summarize the best fit values of $M_G$ and $M_{VF}$ based on a simultaneous fit to all three gauge couplings in Table~\ref{tb:mult}. We set $\alpha_G=0.3$ and consider the scenarios with common superpartner masses at a low scale, $M_{SUSY} = 3$ TeV, and common superpartner masses at the GUT scale,  $M_{SUSY,0} = 9$ TeV.  For each of these models, all three gauge couplings are reproduced at least within $1.5\%$ of their measured values. The GUT scale varies slightly between $3-6\times 10^{16}$ GeV.

\begin{table}[t]
\centering
\begin{tabular}{ |p{3cm}| |c|c||c|c|}
\hline
 			& \multicolumn{2}{|c||}{$M_{SUSY}=3$ TeV}&\multicolumn{2}{|c|}{$M_{SUSY,0}=9$ TeV}\\
\hline
\centering Model& $M_G$ (GeV) & $M_{VF}$ (GeV)&$M_G$ (GeV) &$M_{VF}$ (GeV)\\
\hline
\centering $n_{16}=1$ &$4.06\times 10^{16}$ &970.12 &3.20$\times 10^{16}$ &$886.05$\\
\hline
\centering $n_{16}=2$ &$5.87\times 10^{16}$&$4.04\times 10^{9}$&$3.42\times 10^{16}$ &$2.48\times 10^{9}$\\
\hline
\centering $n_5=4$ &$4.48\times 10^{16}$ & $685.73$&$5.81\times 10^{16}$ & $638.46$\\
\hline
\centering $n_5=5$&$3.86\times 10^{16}$ &$2.15\times 10^5$ &$4.66\times 10^{16}$ & $2.07\times 10^5$\\
\hline
\centering $n_{10}=2$&$2.98\times 10^{16}$&$1.42\times 10^7$ &$3.44\times 10^{16}$ &$1.40\times 10^7$\\
\hline
\centering $n_{10}=3$ &$4.69\times 10^{16}$&$2.11\times 10^{10}$&$3.36\times 10^{16}$ &$1.52\times 10^{10}$\\
\hline
\end{tabular}
\caption{The best fit $M_G$ and $M_{VF}$ in various extensions of the MSSM for $\alpha_G=0.3$ assuming $M_{SUSY} = 3$ TeV and $M_{SUSY,0} = 9$ TeV. }
\label{tb:mult}
\end{table}

\section{Top yukawa coupling}
\label{sec:top}

In this section we investigate whether the top quark mass or its Yukawa coupling can also be understood  from the IR fixed point behavior.
There are two immediate difficulties with this task. First, the MSSM is a two Higgs doublet model and thus the top quark mass is determined not only from the top Yukawa coupling but also from the structure of vacuum expectation values of the two Higgs doublets parametrized by $\tan \beta$. Therefore, even if there is a prediction for the top quark Yukawa coupling, it does not directly translate into a prediction for the top quark mass. Nevertheless, one can instead use the measured top quark mass to predict  $\tan \beta$ or at least conclude that the understanding of the top quark mass from the IR fixed point  is possible.

The second complication is that, in the MSSM+1VF, there can be up to two additional large  Yukawa couplings of $H_u$ to vectorlike quarks (in the basis where Yukawa couplings to $H_u$ are diagonal),
\begin{equation}
W\supset Y_{U}H_uQ\bar{U} + Y_{D}H_u\bar{Q}D,
\end{equation}
which affect the RG flow of the top Yukawa coupling and thus the IR fixed point prediction (we do not consider here Yukawa couplings of $H_u$ to vectorlike leptons since these do not have a large effect). 
The one-loop beta function for the top Yukawa coupling (neglecting the bottom Yukawa coupling) is then given by
\begin{equation}
\beta^{(1)}_{y_t} = y_t\left(6y_t^2 + 3Y_{U}^2 + 3Y_{D}^2 - \frac{16}{3}g_3^2 - 3g_2^2 - \frac{13}{15}g_1^2\right)
\label{eq:b_yt}
\end{equation}
and the two loop beta function including threshold effects from superpartners and vectorlike matter can be found in the Appendix.

Neglecting the additional Yukawa couplings  for the moment, the $\beta$-function vanishes (and the IR fixed point occurs) when
 \begin{equation}
 y_t^2 \simeq \frac{8}{9}g_3^2 + \frac{1}{2}g_2^2. 
 \end{equation}
If the boundary condition for $y_t$ at the GUT scale is above the IR fixed point, the positive contribution from $y_t$ itself dominates the RG evolution and drives $y_t$ down while, for the boundary condition below the IR fixed point, the negative contribution from the gauge couplings dominates and drives $y_t$ up~\cite{Pendleton:1981}. However, in the SM or in the MSSM (if the top quark mass is below the IR fixed point which depends on $\tan\beta$), the IR fixed point behavior is not very effective.  Although the top quark mass happens to be near the predicted IR fixed point, the boundary condition for $y_t$ at the GUT scale is already close to the IR fixed point in both models because the gauge couplings (including $\alpha_3$)  are small  in most of the energy interval between the GUT scale and the EW scale. 

However, the IR fixed point behavior for $y_t$ is  very effective in models with asymptotically divergent couplings because of large  gauge couplings (especially $\alpha_3$) over the whole energy interval~\cite{Dermisek:2012as}. This occurs no matter if the GUT scale boundary condition is far above or far below the IR fixed point. Thus, these models typically have a very sharp prediction for $y_t$ at the EW scale. If the extra matter has significant Yukawa couplings to $H_u$ the  prediction broadens since the large Yukawa couplings in the model  share the IR fixed point value as can be seen from Eq.~(\ref{eq:b_yt}). 
Given that the number and size of Yukawa couplings of vectorlike matter is not known, we will consider scenarios with no additional Yukawa couplings, one and two additional large Yukawa couplings. We will assume a universal boundary condition for all the Yukawas but also discuss the variation of predictions when departing from the universality assumption.

 \begin{figure}[t]
\centering
\includegraphics[width = 3in]{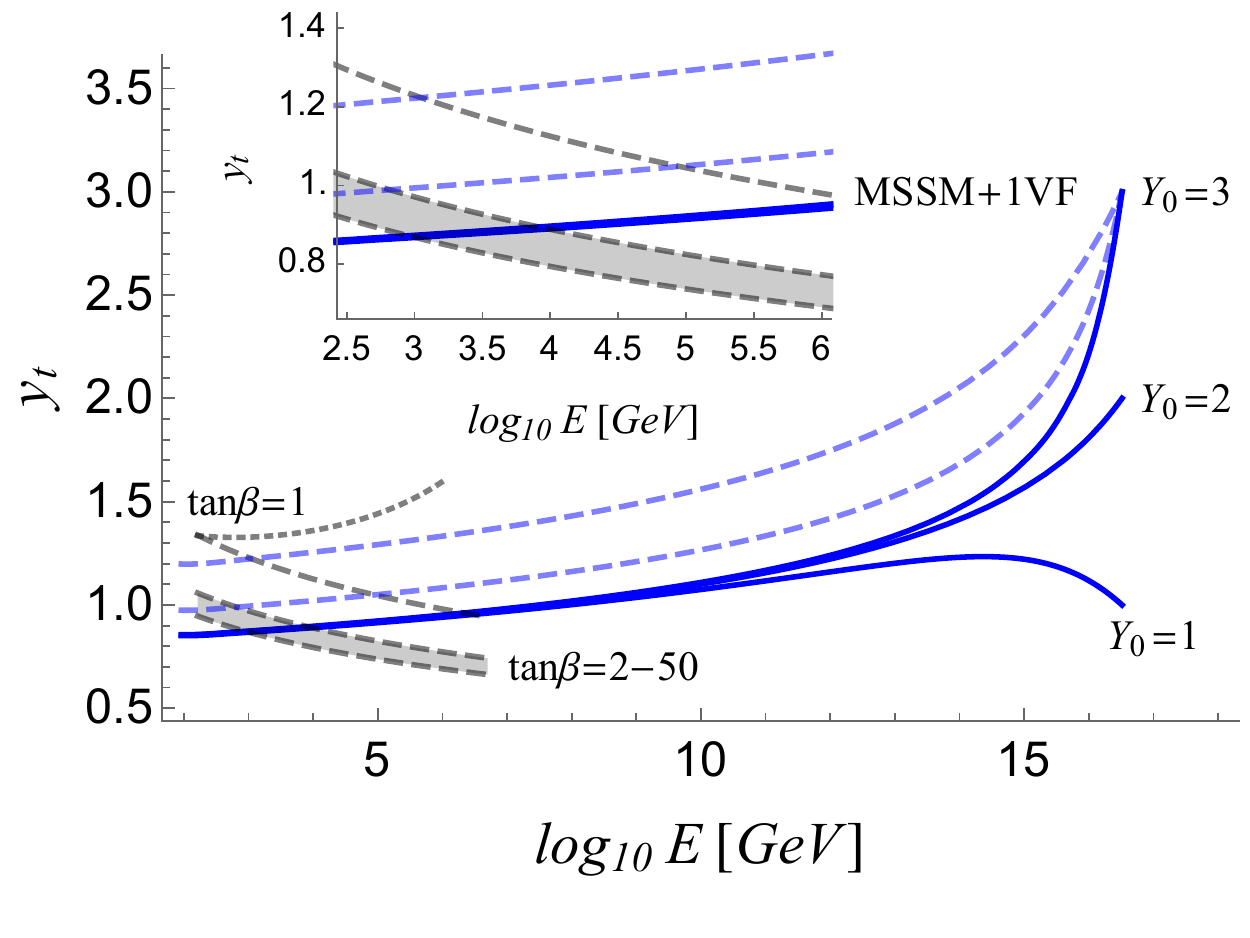}
\includegraphics[width = 3in]{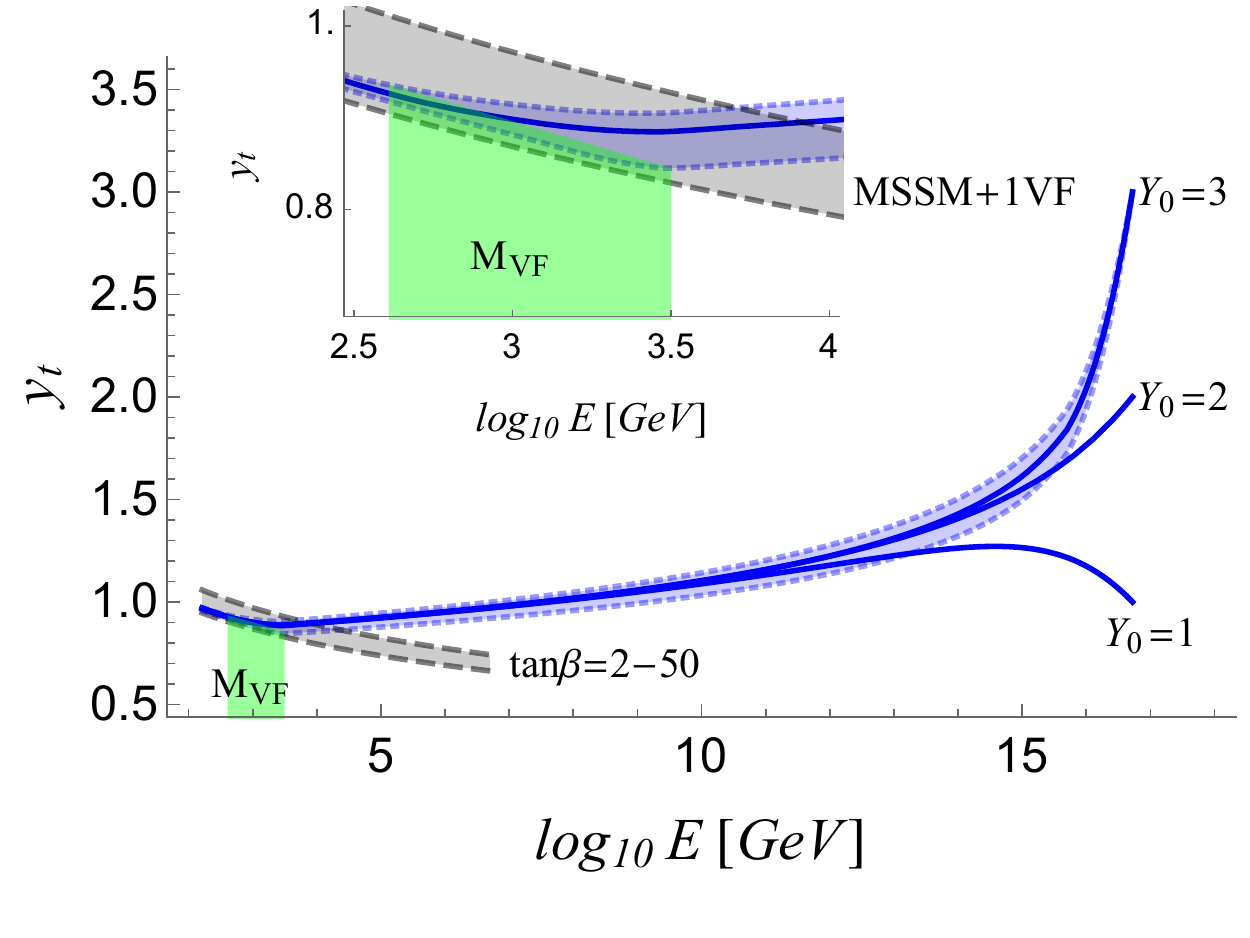}\\
\hspace{1cm}(a)\hspace{7cm} (b)
\caption{(a) RG evolution of $y_t$ in the MSSM+1VF for $\alpha_G = 0.3$ with no additional Yukawa coupling (upper dashed blue), one additional  coupling (lower dashed blue) and two additional  couplings (solid blue) assuming universal boundary condition for all couplings, $Y_0=3$. For the last case the RG evolution is also shown for $Y_0=1$ and 2.  No thresholds from superpartners or vectorlike matter are assumed. The gray dashed lines and shaded region at low energies show the evolution of $y_t$  in the two Higgs doublet model obtained from the measured value of the top quark mass for $\tan\beta=1,$ 2 and 50 assuming that all Higgs bosons except the SM-like one are heavy.  The gray dotted line for $\tan \beta = 1$ assumes that all Higgs bosons are at the EW scale. The inset zooms in the region at low energies.
(b) The same as in (a) for the case with two additional Yukawa couplings but with  $M_{SUSY} = 3$ TeV and the $M_{VF}$ adjusted so that $\alpha_3$ reproduces the measured value. The dashed blue lines and shaded region show the effect of varying $\alpha_G$ between 0.2 and 0.4 for $Y_0=3$. The green highlight shows the range of $M_{VF}$  required by $\alpha_3$ for  $\alpha_G $ between 0.2 and 0.4 with the left edge of the highlighted region corresponding to $\alpha_G = 0.4$ and the  right edge to  $\alpha_G = 0.2$.}
\label{fig:yukawas_run}
\end{figure}

 In Fig.~\ref{fig:yukawas_run}(a), we show the  RG evolution of $y_t$ in the MSSM+1VF for $\alpha_G = 0.3$ with no additional Yukawa couplings (upper dashed blue), one additional Yukawa coupling (lower dashed blue) and two additional Yukawa couplings (solid blue) assuming universal boundary conditions for all couplings, $Y_0=3$.  The RG evolution of the additional couplings (not shown) closely follow the evolution of $y_t$. The full particle content of the MSSM+1VF is assumed all the way to the EW scale. For the  case with two additional couplings, the RG evolution is also shown for $Y_0=1$ and 2. Almost identical EW scale values of  $y_t$ from a large range of boundary conditions at the GUT scale illustrate the advertised effect of approaching the IR fixed point very fast as a result of larger gauge couplings compared to the MSSM. 
 
 To gain an indication of the optimal scale of vectorlike matter and superpartners we also plot the evolution of $y_t$  in the two Higgs doublet model obtained from the measured value of the top quark mass for $\tan\beta=1,$ 2 and 50  assuming that all Higgs bosons except the SM-like one are heavy (gray dashed lines and shaded region at low energies). The coupling is extracted  from the equation $m_t=y_tv\sin\beta$ with  $v=174$ GeV and appropriate corrections from converting the pole mass to the running mass \cite{Pierce:1997}. For $\tan \beta = 1$, we also plot the RG evolution assuming that all Higgs bosons are at the EW scale (gray dotted line). The masses of other Higgs bosons  dramatically  affect the RG evolution of $y_t$ for $\tan \beta = 1$ while for $\tan \beta > 2$ they play only a minor role. A similar line for $\tan \beta = 2$ would be just slightly above the line shown and for $\tan \beta = 50$ the lines would be on top of each other and thus we do not show them. Note also that a line for $\tan \beta = 10$ would not be visibly distinguishable from $\tan \beta = 50$ and thus the whole shaded range effectively corresponds to the variation of $\tan \beta$ between 2 and 10.
   
From the figure we see that obtaining the top quark mass from the IR fixed point in the case with just the top Yukawa coupling requires small $\tan \beta$, light superpartners and light vectorlike matter which is not consistent with experimental limits or constraints from the Higgs boson mass. Thus, for a viable scenario with multi-TeV superpartners and larger $\tan \beta$,   the top Yukawa coupling has to be somewhat below the IR fixed point at the EW scale. Couplings below the IR fixed point are driven to small values at the GUT scale by large gauge couplings. For example, for $\tan \beta = 10$ we need $y_t(M_G) \simeq 0.12$.   Alternatively, the IR fixed point value of the Yukawa coupling can lead to the measured top quark mass if vectorlike masses are not diagonal and the top quark is a mixture of a state with large Yukawa coupling and another one with no Yukawa coupling. In this case a larger Yukawa coupling than naively inferred from the top quark mass is required~\cite{Dermisek:2016tzw}.

Similar comments apply to the case  with one additional Yukawa coupling which also seems to be excluded by the Higgs mass.\footnote{This scenario might be phenomenologically viable assuming lighter vectorlike matter and very heavy superpartners that generate sufficient  Higgs boson mass for small $\tan \beta$. However, such an arrangement is not favored by the results related to understanding of gauge couplings discussed in the previous section.} However, the case with two additional Yukawa couplings points to a multi-TeV  scale for superpartners and vectorlike matter which is not only phenomenologically viable but also simultaneously  favored by understanding the values of the gauge couplings. Thus, in what follows, we will focus on this scenario.

 \begin{table}[t]
\centering
\begin{tabular}{ |p{2cm}||c|c|c|c|}
\hline
\centering$Y_0$&$y_t(m_t)$&$M_{VF}$ (GeV)& $\tan\beta$ for $M_t = 173.1$ GeV & $M_{t}$ (GeV) for $\tan\beta=2-50$\\
\hline
\centering1&$0.979_{-0.004}^{+0.007}$ &$819_{+2909}^{-378}$  &$3.6$& $162.2_{-2.3}^{+1.3}$ - $177.8_{-2.4}^{+0.9}$\\
\hline
\centering  2&$0.968_{-0.003}^{+0.005}$ &$740_{+2543}^{-337}$&$3.7$& $162.0_{-2.2}^{+1.6}$ - $177.5_{-1.1}^{+1.1}$\\
\hline
\centering3 & $0.965_{-0.003}^{+0.004}$ &$720_{+2333}^{-337}$ &$3.9$&$161.4_{-2.4}^{+1.6}$ - $176.8_{-0.9}^{+1.2}$\\
\hline
\end{tabular}
\caption{Predictions for the top Yukawa coupling, $y_t(m_t)$, in the MSSM+1VF for $\alpha_G = 0.3$ and several choices of $Y_0$ assuming two additional  Yukawa couplings to $H_u$. The indicated $M_{VF}$ is required to reproduce the measured value of $\alpha_3(M_Z)$ for $M_{SUSY} = 3$ TeV.  The $\tan \beta$ is the  value that leads to $M_t = 173.1$ GeV and we also show the range of predicted $M_t$ for  $\tan\beta$ between 2 and 50. The superscript and subscript numbers indicate variations resulting from changing  $\alpha_G$ to 0.4 and 0.2 respectively (variations of $\tan\beta$ predictions are negligible). Numerical entries correspond to Fig.~\ref{fig:yukawas_run}(b).}
\label{tb:y_t_tb_10}
\end{table}

 In Fig.~\ref{fig:yukawas_run}(b), we show the impact of integrating out superpartners and vectorlike matter on the IR predictions of $y_t$ in the case with two additional Yukawa couplings. We set $M_{SUSY} = 3$ TeV and adjust $M_{VF}$  so that $\alpha_3$ reproduces the measure value. The dashed blue lines and shaded region show the effect of varying $\alpha_G$ between 0.2 and 0.4, for $Y_0=3$. The green highlight shows the range of $M_{VF}$  required by $\alpha_3$ for  $\alpha_G $ between 0.2 and 0.4 with the left edge of the highlighted region corresponding to $\alpha_G = 0.4$ and the  right edge to  $\alpha_G = 0.2$. The need to integrate out the vectorlike matter at a slightly different scale depending on $\alpha_G$ results in an extra focusing effect at low energies visible in the inset of Fig.~\ref{fig:yukawas_run}(b). The predicted value of  $y_t(m_t)$ is highly insensitive to $\alpha_G$. The numerical values that correspond to Fig.~\ref{fig:yukawas_run}(b) are summarized in Table~\ref{tb:y_t_tb_10} that also contains similar variations of predictions for $Y_0=1,2$ depending on $\alpha_G$ (not shown in  Fig.~\ref{fig:yukawas_run}(b)), the scale of vectorlike matter for all the cases, the $\tan\beta$ required for $M_t = 173.1$ GeV and prediction for  $M_{t}$ for $\tan\beta=2-50$. We see that, assuming universal Yukawa couplings leads to a sharp prediction for  $y_t(m_t)$ that can be translated into a sharp prediction for $\tan\beta$.  The boundary conditions for $\alpha_G$ and $Y_0$ are almost irrelevant.

\begin{figure}[t]
\centering
\includegraphics[width = 3.5in]{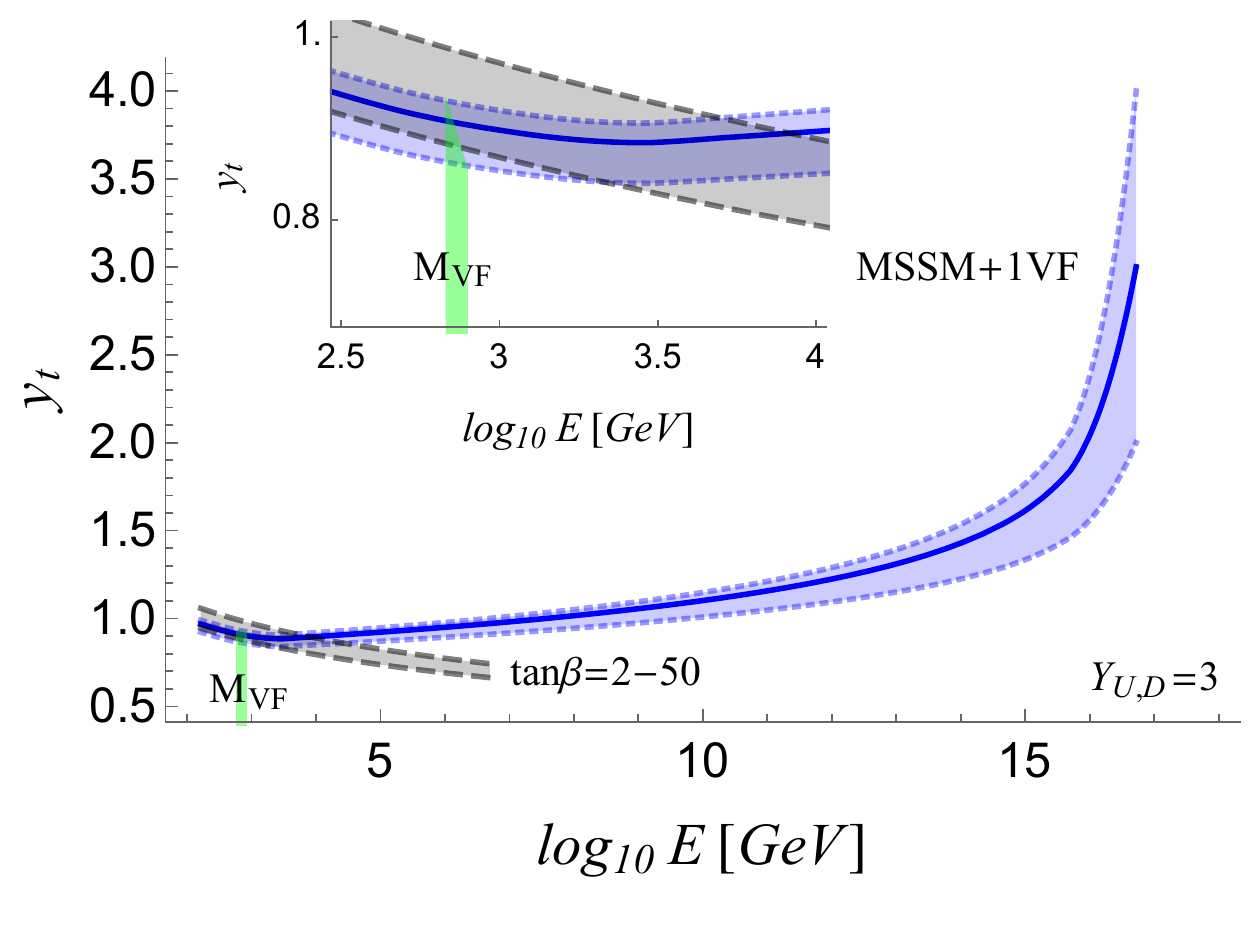}
\caption{RG evolution of $y_t$ in the MSSM+1VF with boundary condition at $M_G$ varied between 2 and 4 for $\alpha_G = 0.3$ and fixed $Y_{U}(M_{G})=Y_{D}(M_{G})=3$. The gray band is the same as in Fig.~\ref{fig:yukawas_run}.
The green highlight shows the range where vectorlike matter is integrated out to obtain the correct $\alpha_3(M_Z)$ with the left edge corresponding to $y_t(M_{G})=4$ and the right edge to $y_t(M_{G})=2$. The inset zooms in the region near $M_{VF}$.}
\label{fig:yt_varied}
\end{figure}

Breaking the universality in large Yukawa couplings expands the region of predicted $\tan \beta$. In Fig.~\ref{fig:yt_varied} we show the RG evolution of $y_t$ in the MSSM+1VF with the boundary condition at $M_G$ varied between 2 and 4 for $\alpha_G = 0.3$ and fixed $Y_{U}(M_{G})=Y_{D}(M_{G})=3$. The green highlight shows the range where vectorlike matter is integrated out to obtain the correct $\alpha_3(M_Z)$ with the left edge corresponding to $y_t(M_{G})=4$ and the right edge to $y_t(M_{G})=2$.  The numerical values that correspond to Fig.~\ref{fig:yt_varied} are summarized in Table~\ref{tb:y_t_vs_YT}.  We see that the correct top quark mass can be obtained for any $\tan \beta > 3$. Although a very sharp prediction is lost,  high insensitivity to all boundary conditions remains.

\begin{table}[t]
\centering
\begin{tabular}{ |p{2cm}| |c|c|c|c|c|}
\hline
\centering$y_t(M_G)$&$y_t(m_t)$&$M_{VF}$ (GeV)&$\tan\beta$ for $M_t = 173.1$ GeV & $M_{t}$ (GeV) for $\tan\beta = 2-50$\\
\hline
\centering2& 0.92&$730$ &-&$154.0-168.8$\\
\hline
\centering  4&1.01  &$690$&3.0&$  165.0- 180.7$\\
\hline
\end{tabular}
\caption{Variations of  predictions quoted in Table~\ref{tb:y_t_tb_10} resulting from breaking the universality in Yukawa couplings. We fix $Y_{U}(M_{G})=Y_{D}(M_{G})=3$ and present results only for $\alpha_G = 0.3$. Numerical entries correspond to Fig.~\ref{fig:yt_varied}.}
\label{tb:y_t_vs_YT}
\end{table}

\section{Conclusions}
\label{sec:conclusions}

We explored extensions  of the MSSM with vectorlike families that feature asymptotically divergent gauge couplings.  In these models,  predictions for gauge and large Yukawa couplings are highly insensitive to boundary conditions at a high scale. 
We used these predictions  to infer the scale of vectorlike matter and superpartners (and  the GUT scale). The results for several extensions of the MSSM are summarized in tables and we discussed in detail  the MSSM extended with one complete vectorlike family, MSSM+1VF ($n_{16} = 1$). This model (together with $n_5=4$ which has identical 1-loop beta functions) stands out since the IR fixed point predictions for the gauge couplings are close  to observed values if vectorlike matter is not far above the EW scale. We considered scenarios with a common mass scale for vectorlike matter (or superpartners) at low energies and also scenarios where vectorlike masses (or superpartners) originated from a universal mass parameter at the GUT scale. 

We find that for any unified gauge coupling, $\alpha_G$, larger than 0.3  vectorlike matter or superpartners   are expected within 1.7 TeV (2.5 TeV) based on all three gauge couplings being simultaneously within 1.5\% (5\%) from observed values. This range  extends to about 4 TeV for $\alpha_G > 0.2$. Increasing masses of superpartners pushes the preferred scale of vectorlike quarks and leptons down and vice versa.  

We have not required that the gauge couplings are reproduced precisely, since significant threshold corrections can originate from superpartner spectrum, spectrum of vectorlike matter or from a specific model at the GUT scale. For example, assuming universal vectorlike masses and universal superpartner masses at a low scale, gauge couplings can be reproduced precisely if GUT scale threshold corrections leading to about $20\%$ splitting of individual couplings at the GUT scale  are assumed.
 This is a very similar result to the usual $3\%$ correction needed in the MSSM, since the GUT scale threshold corrections are proportional to $\alpha_G$ which in our case is about 7 times larger. Alternatively, the gauge couplings can  also be reproduced precisely by splitting individual vectorlike masses within a factor of five between the lightest and heaviest. Interestingly, the spectrum of vectorlike matter resulting from a common mass at the GUT scale leads to much better agreement with measured values of gauge couplings and thus  significantly smaller GUT scale corrections are needed.
 More precise predictions can be made assuming a specific SUSY breaking scenario, specific origin and pattern of vectorlike masses and specific GUT scale model with calculable GUT scale threshold corrections to gauge coupling. Our predictions for the scale of vectorlike matter and superpartners can be considered as central values that can be shifted in both directions. Effects of specific spectrum or GUT scale threshold corrections can be qualitatively inferred from presented plots.

We also find that the IR fixed point behavior for the top Yukawa coupling is  very effective in the MSSM+1VF and there is a very sharp prediction for its value at the EW scale. If the extra matter has significant Yukawa couplings to $H_u$ the  prediction broadens since the large Yukawa couplings in the model  share the IR fixed point value.
In the scenario with two additional large Yukawa couplings of vectorlike quarks the  IR fixed point value of the top Yukawa coupling independently points to a multi-TeV range for vectorlike family and superpartners.  In this scenario, the measured top quark mass can be obtained from the IR fixed point value of the Yukawa coupling for $\tan \beta \simeq 4$ assuming universal value of all large Yukawa couplings at the GUT scale and the range expands to any $\tan \beta > 3$ for significant departures from the universality assumption. 

We have found that the gauge couplings and the top Yukawa coupling can be simultaneously understood  from the IR fixed points in the MSSM+1VF if vectorlike matter is near or somewhat above 1 TeV.
Considering that the Higgs boson mass also points to a multi-TeV range for superpartners in the MSSM, adding a complete vectorlike family at the same scale provides a compelling  scenario where the values of all large couplings in the SM  are understood as a consequence of the particle content of the model.


\vspace{0.2cm}
\noindent
{\bf Acknowledgments:} This work was supported in part by the U.S. Department of Energy under grant number {DE}-SC0010120.

\newpage
\appendix
\section{RG equations for the MSSM with vectorlike matter}

We use  the full set of 2-loop RG equations for extensions of the MSSM with vectorlike matter that we customize to reflect 2-loop threshold corrections to gauge couplings from individual particles in a given model. In addition, we include 3-loop pure gauge terms for the beta functions of the gauge couplings.  For brevity, we define thresholds for SUSY spectra by implicitly summing over generations, e.g. $\theta_{\tilde{q}} = \frac{1}{3}\left(\theta_{\tilde{q}_1} + \theta_{\tilde{q}_2} + \theta_{\tilde{q}_3}\right)$, where $\theta_{\tilde{q}_i}=\theta(\mu-m_{\tilde{q}_i})$ for squarks and similarly for the other superpartners. Subscripts follow the convention that lower case letters are reserved for matter fields of the SM, upper case for additional vectorlike matter, and in either case tildes correspond to respective scalar partners, e.g. for a vectorlike quark and scalar partner we denote $\theta_{Q}$ and $\theta_{\tilde{Q}}$. Gaugino thresholds are given by $\theta_{M_i}$. Higgsinos, $\tilde{H}_u$ and $\tilde{H}_d$ are integrated out together at a scale set by the $\mu$ term and the threshold is denoted $\theta_{\tilde{H}}$. Heavy Higgs contributions are handled similarly with factors of $\theta_{H}$, and we allow contributions for a light SM Higgs to evolve all the way to $M_Z$.

\subsection{Gauge couplings} 
The beta functions for the gauge couplings are given by
\begin{equation}
\frac{d}{dt}g_l = b_l\frac{g_l^3}{16\pi^2} + \frac{g_l^3}{(16\pi^2)^2}\left[\sum_k b_{lk}g_k^2-\sum_{\substack{x=u,d,e\\i = 1,2,3}}C^{x,i}_l(Y_x^{\dagger}Y_x)_{ii} + \sum_{j,k}b_{ljk}\frac{g_j^2g_k^2}{(16\pi^2)}\right],
\end{equation}
where the group theoretical coefficients $b_l, b_{lk}$, and $b_{ljk}$ can be extracted from \cite{Martin:1993}, \cite{Machacek:1983tz}, \cite{Jones:1975}, and \cite{Kolda:1996ea}. Following the procedure described above, we obtain the beta function coefficients with thresholds corresponding to individual particles:
\begin{equation*}
\begin{aligned}
	b_1 =&  \frac{23}{6} + \frac{1}{6}\theta_ t+ \frac{1}{10}\theta_{\tilde{q}} + \frac{3}{10}\theta_{\tilde{l}} + \frac{3}{5}\theta_{\tilde{\bar{e}}} + \frac{4}{5}\theta_{\tilde{\bar{u}}} + \frac{1}{5}\theta_{\tilde{\bar{d}}} + \frac{1}{10}(1 + \theta_{H}) + \frac{2}{5}\theta_{\tilde{H}} \\
 			& + \frac{3}{5}n_5\Big[\frac{4}{9}\theta_{D} + \frac{2}{3}\theta_{L} + \frac{1}{9}\left(\theta_{\tilde{D}} + \theta_{\tilde{\bar{D}}}\right) + \frac{1}{6}\left(\theta_{\tilde{L}} + \theta_{\tilde{\bar{L}}}\right)\Big]\\
			& + \frac{3}{5}n_{10}\Big[\frac{2}{9}\theta_{Q} + \frac{16}{9}\theta_{U} + \frac{4}{3}\theta_{E} + \frac{1}{18}\left(\theta_{\tilde{Q}} + \theta_{\tilde{\bar{Q}}}\right) + \frac{4}{9}\left(\theta_{\tilde{U}} +  \theta_{\tilde{\bar{U}}}\right) + \frac{1}{3}\left(\theta_{\tilde{E}} +  \theta_{\tilde{\bar{E}}}\right)\Big],\\
	b_2 =& -\frac{11}{3} + \frac{1}{3}\theta_t + \frac{4}{3}\theta_{M_{2}} + \frac{3}{2}\theta_{\tilde{q}} + \frac{1}{2}\theta_{\tilde{l}} + \frac{1}{6}(1+\theta_{H}) + \frac{2}{3}\theta_{\tilde{H}}\\
		& + n_5\Big[\frac{2}{3}\theta_{L} + \frac{1}{6}\left(\theta_{\tilde{L}} +  \theta_{\tilde{\bar{L}}}\right)\Big]\\
		& + 3n_{10}\Big[\frac{2}{3}\theta_{Q} + \frac{1}{6}\left(\theta_{\tilde{Q}} +  \theta_{\tilde{\bar{Q}}}\right)\Big],\\
\end{aligned}		
\end{equation*}
\begin{equation}
\begin{aligned}
	b_3 =& -\frac{23}{3} + \frac{2}{3}\theta_t + 2\theta_{M_{3}}  + \theta_{\tilde{q}} + \frac{1}{2}\theta_{\tilde{\bar{u}}}  + \frac{1}{2}\theta_{\tilde{\bar{d}}}\\
		& + n_5 \Big[\frac{2}{3}\theta_{D} + \frac{1}{6}\left(\theta_{\tilde{D}} +  \theta_{\tilde{\bar{D}}}\right)\Big]
		 + n_{10}\Big[\frac{4}{3}\theta_{Q} + \frac{1}{3}\left(\theta_{\tilde{Q}} +  \theta_{\tilde{\bar{Q}}}\right)
		 + \frac{2}{3}\theta_{U} + \frac{1}{6}\left(\theta_{\tilde{U}} +  \theta_{\tilde{\bar{U}}}\right)\Big],\\%
\end{aligned}
\end{equation}
\begin{equation*}
\begin{aligned}
b_{11} = &\frac{6823}{1800} + \frac{17}{1800}\theta_t + \frac{27}{100}\theta_{\tilde{l}}\Big[2 - \theta_{M_1}\Big]  + \frac{54}{25}\theta_{\tilde{\bar{e}}}\Big[2 - \theta_{M_1}\Big] + \theta_{\tilde{q}}\Big[\frac{1}{50} - \theta_{M_1}\left(\frac{1}{150} + \frac{1}{300}\theta_t\right)\Big]\\	
		       & +\theta_{\tilde{\bar{u}}}\Big[\frac{192}{75} - \theta_{M_1}\left(\frac{64}{75} + \frac{32}{75}\theta_t\right)\Big] + \frac{2}{25}\theta_{\tilde{\bar{d}}}\Big[2 - \theta_{M_1}\Big] + \frac{9}{100}\Big[2\theta_{\tilde{H}} + (1+\theta_H)(2-\theta_{\tilde{H}}\theta_{M_1})\Big]\\
			& + \frac{9}{25}n_5\Big[\frac{1}{2}\theta_{L} + \frac{1}{4}\left(\theta_{\tilde{L}} +  \theta_{\tilde{\bar{L}}}\right)\left(2 - \theta_{M_1}\right) + \frac{4}{27}\theta_{D} + \frac{2}{27}\left(\theta_{\tilde{D}} +  \theta_{\tilde{\bar{D}}}\right)\left(2 - \theta_{M_1}\right)\Big]\\
			& + \frac{9}{25}n_{10}\Big[\frac{1}{54}\theta_{Q} + \frac{1}{108}\left(\theta_{\tilde{Q}} +  \theta_{\tilde{\bar{Q}}}\right)\left(2 - \theta_{M_1}\right) + \frac{64}{27}\theta_{U} + \frac{64}{54}\left(\theta_{\tilde{U}} +  \theta_{\tilde{\bar{U}}}\right)\left(2 - \theta_{M_1}\right)\\
			& + 4\theta_{E} + 2\left(\theta_{\tilde{E}} +  \theta_{\tilde{\bar{E}}}\right)\left(2 - \theta_{M_1}\right)\Big],\\
b_{12}  =& \frac{71}{40} + \frac{1}{40}\theta_t+ \frac{27}{20}\theta_{\tilde{l}}\Big[2 - \theta_{M_2}\Big] 
		+ \theta_{\tilde{q}}\Big[\frac{9}{10} - \theta_{M_2}\left(\frac{3}{10} + \frac{3}{20}\theta_t\right)\Big]\\
		&+\frac{9}{20}\Big[2\theta_{\tilde{H}} + (1+\theta_H)(2-\theta_{\tilde{H}}\theta_{M_1})\Big]\\
		& + \frac{9}{10}n_5\Big[\theta_{L} + \frac{1}{2}\left(\theta_{\tilde{L}} +  \theta_{\tilde{\bar{L}}}\right)\left(2 - \theta_{M_1}\right)\Big]\\
		& + \frac{3}{10}n_{10}\Big[\theta_{Q} + \frac{1}{2}\left(\theta_{\tilde{Q}} +  \theta_{\tilde{\bar{Q}}}\right)\left(2 - \theta_{M_1}\right)\Big], \\
b_{13} =& \frac{386}{45} + \frac{10}{45}\theta_t + \theta_{\tilde{q}}\Big[\frac{24}{15} - \theta_{M_3}\left(\frac{8}{15} + \frac{4}{15}\theta_t\right)\Big]
                    + \theta_{\tilde{\bar{u}}}\Big[\frac{192}{15} - \theta_{M_3}\left(\frac{64}{15} + \frac{32}{15}\theta_t\right)\Big]\\
                    &+ \frac{8}{5}\theta_{\tilde{\bar{d}}}\Big[2 - \theta_{M_3}\Big] + \frac{16}{15}n_5\Big[\theta_{D} + \frac{1}{2}\left(\theta_{\tilde{D}} +  \theta_{\tilde{\bar{D}}}\right)\left(2 - \theta_{\tilde{g}}\right)\Big]\\
		& + \frac{8}{15}n_{10}\Big[\theta_{Q} + \frac{1}{2}\left(\theta_{\tilde{Q}} +  \theta_{\tilde{\bar{Q}}}\right)\left(2 - \theta_{\tilde{g}}\right) + 8\theta_{U} + 4\left(\theta_{\tilde{U}} +  \theta_{\tilde{\bar{U}}}\right)\left(2 - \theta_{\tilde{g}}\right)\Big],\\
b_{21} =& \frac{7}{12} + \frac{1}{60}\theta_t + \theta_{\tilde{q}}\Big[\frac{3}{10} - \theta_{M_1}\left(\frac{1}{10} + \frac{1}{20}\theta_t\right)\Big] +\frac{3}{20}\Big[2\theta_{\tilde{H}} + (1+\theta_H)(2-\theta_{\tilde{h}}\theta_{M_1})\Big]\\
		&+\frac{3}{10}n_5\Big[\theta_{L} + \frac{1}{2}\left(\theta_{\tilde{L}} +  \theta_{\tilde{\bar{L}}}\right)\left(2 - \theta_{M_1}\right)\Big] +\frac{1}{10}n_{10}\Big[\theta_{Q} + \frac{1}{2}\left(\theta_{\tilde{Q}} +  \theta_{\tilde{\bar{Q}}}\right)\left(2 - \theta_{M_1}\right)\Big],\\
b_{22} =& -\frac{5}{12}  + \frac{49}{12}\theta_t + \frac{64}{3}\theta_{M_2} + \theta_{\tilde{l}}\Big[\frac{13}{2} - \frac{33}{4}\theta_{M_2}\Big] + \theta_{\tilde{q}}\Big[\frac{39}{2} - \theta_{M_2}\left(\frac{33}{2} + \frac{33}{4}\theta_t\right)\Big]\\ 
		&+\frac{49}{6}\theta_{\tilde{H}} + (1+\theta_{H})(\frac{13}{6}-\frac{11}{4}\theta_{\tilde{H}}\theta_{M_2})\\
		& + n_5\Big[\frac{49}{6}\theta_{L} + \left(\theta_{\tilde{L}} + \theta_{\tilde{\bar{L}}}\right)\left(\frac{13}{6} - \frac{11}{4}\theta_{M_2}\right)\Big] + n_{10}\Big[\frac{49}{2}\theta_{Q} + \left(\theta_{\tilde{Q}} + \theta_{\tilde{\bar{Q}}}\right)\left(\frac{13}{2} - \frac{33}{4}\theta_{M_2}\right)\Big],\\	
b_{23} =& 8 + 4\theta_t +  \theta_{\tilde{q}}\Big[24 - \theta_{M_3}\left(8 + 4\theta_t\right)\Big] + 8n_{10}\Big[\theta_{Q} + \frac{1}{2}\left(\theta_{\tilde{Q}} +  \theta_{\tilde{\bar{Q}}}\right)\left(2 - \theta_{M_3}\right)\Big],\\
\end{aligned}
\end{equation*}	
\begin{equation*}
\begin{aligned}
b_{31} =& \frac{49}{60} + \frac{17}{60}\theta_t+ \theta_{\tilde{q}}\Big[\frac{1}{5} - \theta_{M_1}\left(\frac{1}{15} + \frac{1}{30}\theta_t\right)\Big] + \theta_{\tilde{\bar{u}}}\Big[\frac{24}{15} - \theta_{M_1}\left(\frac{8}{15} + \frac{4}{15}\theta_t\right)\Big] + \frac{1}{5}\theta_{\tilde{\bar{d}}}\Big[2 - \theta_{M_1}\Big]\\
		& + \frac{2}{15}n_5\Big[\theta_{D} +\frac{1}{2}\left(\theta_{\tilde{D}} +  \theta_{\tilde{\bar{D}}}\right)\left(2 - \theta_{M_1}\right)\Big]\\
		& + \frac{1}{15}n_{10}\Big[\theta_{Q} +\frac{1}{2}\left(\theta_{\tilde{Q}} +  \theta_{\tilde{\bar{Q}}}\right)\left(2 - \theta_{M_1}\right)+ 8\theta_{U} + 4\left(\theta_{\tilde{U}} +  \theta_{\tilde{\bar{U}}}\right)\left(2 - \theta_{M_1}\right)\Big],\\
\end{aligned}
\end{equation*}
\begin{equation}
\begin{aligned}
b_{32} =& \frac{15}{4} + \frac{3}{4}\theta_t + \theta_{\tilde{q}}\Big[9 - \theta_{M_2}\left(3 + \frac{3}{2}\theta_t\right)\Big] + 3n_{10}\Big[\theta_{Q} + \frac{1}{2}\left(\theta_{\tilde{Q}} +  \theta_{\tilde{\bar{Q}}}\right)\left(2 - \theta_{M_2}\right)\Big],\\
b_{33} =& -\frac{116}{3} + \frac{38}{3}\theta_t + 48\theta_{M_3} + \theta_{\tilde{\bar{d}}}\Big(11 - 13\theta_{M_3}\Big) + \theta_{\tilde{q}}\Big[22 - \theta_{M_3}\left(\frac{52}{3} + \frac{26}{3}\theta_t\right)\Big]\\
 		&+ \theta_{\tilde{\bar{u}}}\Big[11 - \theta_{M_3}\left(\frac{26}{3} + \frac{13}{3}\theta_t\right)\Big]+ n_5\Big[\frac{38}{3}\theta_{D} + \left(\theta_{\tilde{Q}} + \theta_{\tilde{\bar{Q}}}\right)\left(\frac{11}{3} - \frac{13}{3}\theta_{M_3}\right)\Big]\\
  		&+n_{10}\Big[\frac{76}{3}\theta_{Q} + \left(\theta_{\tilde{Q}} + \theta_{\tilde{\bar{Q}}}\right)\left(\frac{22}{3} - \frac{26}{3}\theta_{M_3}\right) + \frac{38}{3}\theta_{U} + \left(\theta_{\tilde{U}} + \theta_{\tilde{\bar{U}}}\right)\left(\frac{11}{3} - \frac{13}{3}\theta_{M_3}\right)\Big].
\end{aligned}
\end{equation}
The matrix appearing in the 2-loop contribution from Yukawa couplings is given by
\begin{equation}
C_l^{x,i} = \begin{pmatrix}
					\frac{17}{10} + (\theta_{\tilde{q}_i}+ \frac{25}{10}\theta_{\tilde{u}_i})\theta_{\tilde{H}}&\frac{1}{2}+ (\theta_{\tilde{q}_i} + \frac{13}{10}\theta_{\tilde{d}_i})\theta_{\tilde{H}}&\frac{3}{2} +( \frac{3}{5}\theta_{\tilde{l}_i}									 + \frac{3}{2}\theta_{\tilde{e}_i})\theta_{\tilde{H}}\\
					    \frac{3}{2} + 3(\theta_{\tilde{q}_i} + \frac{1}{2}\theta_{\tilde{u}_i})\theta_{\tilde{H}}& \frac{3}{2} + 3(\theta_{\tilde{q}_i} + \frac{1}{2}\theta_{\tilde{d}_i})\theta_{\tilde{H}}&\frac{1}{2} + (\theta_{\tilde{l}_i} 
					    	+ \frac{1}{2}\theta_{\tilde{e}_i})\theta_{\tilde{H}}\\
					    2+ (\theta_{\tilde{q}_i} + \theta_{\tilde{u}_i})\theta_{\tilde{H}}&2+ (\theta_{\tilde{q}_i} + \theta_{\tilde{d}_i})\theta_{\tilde{H}}&0	
			\end{pmatrix},
\end{equation}
where $x= u,d,e$ for up, down, and lepton Yukawa couplings and $i = 1, 2, 3$ for each generation.\\
\noindent
For the 3-loop pure gauge contributions to the beta functions for gauge couplings we integrate out the SUSY spectrum at a common scale with $\theta_{SUSY}$ and similarly for the contributions from vectorlike matter with $\theta_{n5}$ and $\theta_{n_{10}}$:
\vspace{0.5cm}
\begin{equation*}
\begin{aligned}
b_{111} &= -\frac{194293}{12000} - \frac{277817}{4000}\theta_{SUSY} - \frac{7507}{450}n_5\theta_{n_5} - \frac{12859}{150}n_{10}\theta_{n_{10}} - \frac{7}{10}n_5^2\theta_{n_5} - \frac{207}{10}n_{10}^2\theta_{n_{10}} - 9n_5n_{10}\theta_{n_5}\theta_{n_{10}},\\
b_{112} &= \frac{123}{160} - \frac{2823}{800}\theta_{SUSY} - \frac{27}{25}n_5\theta_{n_5} - \frac{1}{25}n_{10}\theta_{n_{10}},\\
b_{113} &= -\frac{137}{75} + \frac{959}{75}\theta_{SUSY} - \frac{128}{225}n_5\theta_{n_5} - \frac{688}{75}n_{10}\theta_{n_{10}},\\
b_{121} &= 0,\\
b_{122} &= \frac{789}{64} - \frac{9129}{320}\theta_{SUSY} - \frac{27}{2}n_5\theta_{n_5} - \frac{261}{10}n_{10}\theta_{n_{10}} - \frac{27}{10}n_5^2\theta_{n_5} - \frac{27}{10}n_{10}^2\theta_{n_{10}} - 9n_5n_{10}\theta_{n_5}\theta_{n_{10}},\\
b_{123} &= -\frac{3}{5} - \frac{21}{5}\theta_{SUSY} - \frac{16}{5}n_{10}\theta_{n_{10}},\\
\end{aligned}
\end{equation*}
\begin{equation*}
\begin{aligned}
b_{131} &= 0,\\
b_{132} &= 0,\\
b_{133} &= \frac{297}{5} - \frac{407}{15}\theta_{SUSY} - \frac{1012}{45}n_5\theta_{n_5} - \frac{308}{5}n_{10}\theta_{n_{10}} - \frac{16}{5}n_5^2\theta_{n_5} - \frac{216}{5}n_{10}^2\theta_{n_{10}} - 24n_5n_{10}\theta_{n_5}\theta_{n_{10}},\\
b_{211} &= -\frac{10077}{1600} - \frac{19171}{1600}\theta_{SUSY} - \frac{441}{50}n_5\theta_{n_5} - \frac{1513}{150}n_{10}\theta_{n_{10}} - \frac{9}{10}n_5^2\theta_{n_5} - \frac{9}{10}n_{10}^2\theta_{n_{10}} - \frac{12}{5}n_5n_{10}\theta_{n_5}\theta_{n_{10}},\\
b_{212} &= \frac{873}{160} - \frac{117}{32}\theta_{SUSY} + \frac{3}{5}n_5\theta_{n_5} + \frac{1}{5}n_{10}\theta_{n_{10}},\\
b_{213} &= -\frac{1}{5} - \frac{7}{5}\theta_{SUSY} - \frac{16}{15}n_{10}\theta_{n_{10}},\\
b_{221} &= 0,\\
b_{222} &= \frac{324953}{1728} - \frac{264473}{1728}\theta_{SUSY} - \frac{33}{2}n_5\theta_{n_5} - \frac{99}{2}n_{10}\theta_{n_{10}} - \frac{13}{2}n_5^2\theta_{n_5} - \frac{117}{2}n_{10}^2\theta_{n_{10}} - 39n_5n_{10}\theta_{n_5}\theta_{n_{10}},\\
b_{223} &= 39 - 15\theta_{SUSY} + 16n_{10},\\
b_{231} &= 0,\\
b_{232} &= 0,\\
b_{233} &= 81 - 37\theta_{SUSY} - 36n_5\theta_{n_5} - \frac{236}{3}n_{10}\theta_{n_{10}} - 72n_{10}^2\theta_{n_{10}} - 24n_5n_{10}\theta_{n_5}\theta_{n_{10}},\\
b_{311} &= -\frac{523}{120} - \frac{3667}{200}\theta_{SUSY} - \frac{2689}{450}n_5\theta_{n_5} - \frac{3353}{150}n_{10}\theta_{n_{10}} - \frac{2}{5}n_5^2\theta_{n_5} - \frac{27}{5}n_{10}^2\theta_{n_{10}} - 3n_5n_{10}\theta_{n_5}\theta_{n_{10}},\\
b_{312} &= -\frac{3}{40} - \frac{21}{40}\theta_{SUSY} - \frac{2}{5}n_{10}\theta_{n_{10}},\\
b_{313} &= \frac{77}{15} - \frac{11}{3}\theta_{SUSY} + \frac{8}{45}n_5\theta_{n_5} + \frac{4}{5}n_{10}\theta_{n_{10}},\\
b_{321} &= 0,\\
\end{aligned}
\end{equation*}
\begin{equation}
\begin{aligned}
b_{322} &= \frac{109}{8} - \frac{325}{8}\theta_{SUSY} - \frac{27}{2}n_5\theta_{n_5} - \frac{117}{2}n_{10}\theta_{n_{10}} - 27n_{10}^2\theta_{n_{10}} - 9n_5n_{10}\theta_{n_5}\theta_{n_{10}},\\
b_{323} &= 21 - 15\theta_{SUSY} + 4n_{10}\theta_{n_{10}},\\
b_{331} &= 0,\\
b_{332} &= 0,\\
b_{333} &= \frac{65}{2} + \frac{499}{6}\theta_{SUSY} + \frac{430}{9}n_5\theta_{n_5} + \frac{430}{3}n_{10}\theta_{n_{10}} - 11n_5^2\theta_{n_5} - 99n_{10}^2\theta_{n_{10}} - 66n_5n_{10}\theta_{n_5}\theta_{n_{10}}.
\end{aligned}
\end{equation}

\subsection{Vectorlike mass terms}

The fermion mass terms of  vectorlike fields originate from the superpotential 
\begin{equation}
W\supset M_{Q}Q\bar{Q} + M_{U}U\bar{U} + M_{E}E\bar{E} + M_{L}L\bar{L} + M_{N}N\bar{N} + M_{D}D\bar{D},
\end{equation}
where the fields transform under $SU(3)_C \times SU(2)_L \times U(1)_Y$ as 
\begin{equation}
\begin{aligned}
Q&=(3,2,\frac{1}{6}), \hspace{0.5cm} \bar{Q}=(\bar{3},2,-\frac{1}{6}), \hspace{0.5cm} U=(3,1,\frac{2}{3}), \hspace{0.5cm} \bar{U}=(\bar{3},1,-\frac{2}{3}),\\
D&=(3,1,-\frac{1}{3}), \hspace{0.5cm} \bar{D}=(\bar{3},1,\frac{1}{3}), \hspace{0.5cm} L=(1,2,-\frac{1}{2}), \hspace{0.5cm} \bar{L}=(1,2,\frac{1}{2}),\\
E&=(1,1,-1), \hspace{0.5cm} \bar{E}=(1,1,1), \hspace{0.5cm} N=(1,1,0), \hspace{0.5cm} \bar{N}=(1,1,0).
\end{aligned}
\end{equation}
The RG equation for vectorlike fermion masses  can be obtained in a similar way as that of the $\mu$-term in the MSSM,
\begin{equation}
\frac{dM_V}{dt} = M_V\Big[\frac{1}{16\pi^2}\Gamma^{(1)}_V + \frac{1}{(16\pi^2)^2}\Gamma^{(2)}_V\Big]
\end{equation}
with $\Gamma^{(i)}_V = \gamma^{(i)}_V + \gamma^{(i)}_{\bar{V}}$ and $V=Q,U,E,L,N,D$. Here we include two-loop RG equations with one-loop threshold corrections from individual fields. Neglecting Yukawa couplings, the anomalous dimensions of the fields and their conjugates are identical and we obtain (note, in this approximation $\Gamma^{(1)}_N$, $\Gamma^{(2)}_N =0$):\\
\begin{equation}
\begin{aligned}
\Gamma^{(1)}_Q &= -\frac{1}{30}g_1^2\left[3 - \frac{1}{2}(\theta_{\tilde{Q}^V} + \theta_{\tilde{\bar{Q}}^V})\theta_{M_1}\right] - \frac{3}{2}g_2^2\left[(3 - \frac{1}{2}(\theta_{\tilde{Q}^V} + \theta_{\tilde{\bar{Q}}^V})\theta_{M_2}\right] - \frac{8}{3}g_3^2\left[3 - \frac{1}{2}(\theta_{\tilde{Q}^V} + \theta_{\tilde{\bar{Q}}^V})\theta_{M_3}\right],\\
\Gamma^{(2)}_Q &= \frac{319}{450}g_1^4 + \frac{39}{2}g_2^4 + \frac{176}{9}g_3^4 + \frac{1}{5}g_1^2g_2^2 + \frac{16}{45}g_1^2g_3^2 + 16g_2^2g_3^2,\\
\Gamma^{(1)}_U &= -\frac{8}{15}g_1^2\left[3 - \frac{1}{2}(\theta_{\tilde{U}^V} + \theta_{\tilde{\bar{U}}^V})\theta_{M_2})\right]- \frac{8}{3}g_3^2\left[3 -\frac{1}{2}(\theta_{\tilde{U}^V} + \theta_{\tilde{\bar{U}}^V})\theta_{M_3}\right],\\
\Gamma^{(2)}_U &= \frac{2672}{225}g_1^4 + \frac{176}{9}g_3^4 + \frac{256}{45}g_1^2g_3^2,\\
\Gamma^{(1)}_E &= -\frac{6}{5}g_1^2\left[3 -\frac{1}{2}(\theta_{\tilde{E}^V} + \theta_{\tilde{\bar{E}}^V})\theta_{M_1}\right],\\
\Gamma^{(2)}_E &= \frac{708}{25}g_1^4,\\
\Gamma^{(1)}_L &= -\frac{3}{10}g_1^2\left[3 -\frac{1}{2}(\theta_{\tilde{L}^V} + \theta_{\tilde{\bar{L}}^V})\theta_{M_1}\right] - \frac{3}{2}g_2^2\left[3 - \frac{1}{2}(\theta_{\tilde{L}^V} + \theta_{\tilde{\bar{L}}^V})\theta_{M_2}\right],\\
\Gamma^{(2)}_L &= \frac{327}{50}g_4 + \frac{39}{2}g_2^4 + \frac{9}{5}g_1^2g_2^2,\\
\Gamma^{(1)}_D &= -\frac{2}{15}g_1^2\left[3 -\frac{1}{2}(\theta_{\tilde{D}^V} + \theta_{\tilde{\bar{D}}^V})\theta_{M_1}\right] - \frac{8}{3}g_3^2\left[3 - \frac{1}{2}(\theta_{\tilde{D}^V} + \theta_{\tilde{\bar{D}}^V})\theta_{M_3}\right],\\
\Gamma^{(2)}_D &= \frac{644}{225}g_1^4 + \frac{176}{9}g_3^4 + \frac{64}{45}g_1^2g_3^2.
\end{aligned}
\end{equation}
In general, there will also be soft-mass terms corresponding to the scalar partners for vectorlike matter. We do not list their RGEs here as they are quite cumbersome and almost exactly the same as those in the MSSM differing only by a minus sign on terms that sum over the hypercharge generator for conjugate fields \cite{Martin:1993}.

\subsection{Top Yukawa coupling and Yukawa couplings of vectorlike fields}

In Sec.~\ref{sec:top}, we introduced additional Yukawa couplings from vectorlike fields coupling to $H_u$,
\begin{equation}
W\supset Y_{U}H_uQ\bar{U} + Y_{D}H_u\bar{Q}D,
\end{equation}
that will modify the RG evolution of the top Yukawa coupling. The beta function for the top Yukawa coupling,
\begin{equation}
\frac{1}{y_t}\frac{dy_t}{dt}=\frac{1}{16\pi^2}\beta^{(1)}_{y_t} + \frac{1}{(16\pi^2)^2}\beta^{(2)}_{y_t},
\end{equation}
can be extracted from~\cite{Martin:1993, Machacek:1983fi}.With the additional Yukawa couplings we have
\begin{equation}
\beta_{y_t}^{(1)} = 6y_t^2 + 3Y_U^2 + 3Y_D^2 - \frac{16}{3}g_3^2 - 3g_2^2 - \frac{13}{15}g_1^2,
\end{equation}
at one loop. For the two loop beta function we find
\begin{equation}
\begin{aligned}
\beta_{y_t}^{(2)} =& -22y_t^4 - 9Y_U^4 - 9Y_D^4 - 9y_t^2(Y_U^2 + Y_D^2) + (16g_3^2 + 6g_2^2 + \frac{6}{5}g_1^2)y_t^2\\
				& +16g_3^2(Y_U^2 + Y_D^2) + \frac{2}{5}g_1^2(2Y_U^2 - Y_D^2) + \frac{4303}{450}g_1^4 + \frac{39}{2}g_2^4 + \frac{176}{9}g_3^4\\
				&+g_2^2g_1^2 + 8g_3^2g_2^2 + \frac{136}{45}g_3^2g_1^2.
 \end{aligned}
\end{equation}
The beta function for $Y_{U}$ is identical to that for $y_t$ and the beta function for $Y_{D}$ differs only through terms proportional to hypercharge factors. Yukawa couplings from vector-like matter stop contributing to the RG flow at $M_{VF}$. At the SUSY scale we match to the RG evolution in the SM. In our analysis we have also included corrections from switching between $\overline{\text{DR}}$ and $\overline{\text{MS}}$ schemes following the recipe of \cite{Martin:1993yx}.




\begin{thebibliography}{99}

\bibitem{PDG_GUTs} For a review and references, see the section on grand unified theories in Ref.~\cite{Patrignani:2016xqp}.
 
 
\bibitem{Patrignani:2016xqp} 
  C.~Patrignani {\it et al.} [Particle Data Group],
  Chin.\ Phys.\ C {\bf 40}, no. 10, 100001 (2016).


\bibitem{Maiani:1977cg} 
  L.~Maiani, G.~Parisi and R.~Petronzio,
  Nucl.\ Phys.\ B {\bf 136}, 115 (1978).

\bibitem{Cabibbo:1982hy} 
  N.~Cabibbo and G.~R.~Farrar,
  Phys.\ Lett.\ B {\bf 110}, 107 (1982).
  
  
\bibitem{Moroi:1993} 
 T.~Moroi, H.~Murayama, and T.~Yanagida,
  Phys.\ Rev.\ D {\bf 48}, 2995 (1993)
  [hep-ph/9306268].
  

\bibitem{Carone:2017ubr} 
  C.~D.~Carone, S.~Chaurasia and J.~C.~Donahue,
  Phys.\ Rev.\ D {\bf 96}, no. 3, 035002 (2017)
  [arXiv:1705.09716 [hep-ph]].





\bibitem{Dermisek:2012as} 
  R.~Dermisek,
  Phys.\ Lett.\ B {\bf 713}, 469 (2012)
  [arXiv:1204.6533 [hep-ph]].
 
\bibitem{Dermisek:2012ke} 
  R.~Dermisek,
  Phys.\ Rev.\ D {\bf 87}, no. 5, 055008 (2013)
  [arXiv:1212.3035 [hep-ph]].
  
  

\bibitem{Pendleton:1981} 
  B.~Pendleton and G.~G.~Ross,
  Phys.\ Lett.\ B {\bf 98}, 291 (1981).
  





\bibitem{Dermisek:2016zvl} 
  R.~Dermisek,
  arXiv:1611.03188 [hep-ph].
  
  
\bibitem{Dermisek:2017xmd} 
  R.~Dermisek and N.~McGinnis,
  arXiv:1705.01910 [hep-ph].


  
  
  
\bibitem{Babu:1996zv} 
  K.~S.~Babu and J.~C.~Pati,
  Phys.\ Lett.\ B {\bf 384}, 140 (1996).


  
\bibitem{Kolda:1996ea} 
  C.~F.~Kolda and J.~March-Russell,
  Phys.\ Rev.\ D {\bf 55}, 4252 (1997)
  [hep-ph/9609480].
  
\bibitem{Ghilencea:1997yr} 
  D.~Ghilencea, M.~Lanzagorta and G.~G.~Ross,
  Phys.\ Lett.\ B {\bf 415}, 253 (1997)
  [hep-ph/9707462].
  
  
\bibitem{AmelinoCamelia:1998tm} 
  G.~Amelino-Camelia, D.~Ghilencea and G.~G.~Ross,
  Nucl.\ Phys.\ B {\bf 528}, 35 (1998)
  [hep-ph/9804437].
  
  
    
\bibitem{BasteroGil:1999dx} 
  M.~Bastero-Gil and B.~Brahmachari,
  Nucl.\ Phys.\ B {\bf 575}, 35 (2000)
  [hep-ph/9907318].
  

  
\bibitem{Babu:2008ge} 
  K.~S.~Babu, I.~Gogoladze, M.~U.~Rehman and Q.~Shafi,
  Phys.\ Rev.\ D {\bf 78}, 055017 (2008)
  [arXiv:0807.3055 [hep-ph]].
  
  
  
  
\bibitem{Martin:2009bg} 
S.~P.~Martin,
  Phys.\ Rev.\ D {\bf 81}, 035004 (2010).
  
  
\bibitem{Dermisek:2016tzw} 
  R.~Dermisek,
  Phys.\ Rev.\ D {\bf 95}, no. 1, 015002 (2017)
  [arXiv:1606.09031 [hep-ph]].
  
  
    

\bibitem{Choudhury:2001hs}
  D.~Choudhury, T.~M.~P.~Tait and C.~E.~M.~Wagner,
  Phys.\ Rev.\  D {\bf 65}, 053002 (2002)
  [arXiv:hep-ph/0109097].


\bibitem{Dermisek:2011xu} 
  R.~Dermisek, S.~-G.~Kim and A.~Raval,
  Phys.\ Rev.\ D {\bf 84}, 035006 (2011)
  [arXiv:1105.0773 [hep-ph]].


\bibitem{Dermisek:2012qx} 
  R.~Dermisek, S.~-G.~Kim and A.~Raval,
  Phys.\ Rev.\ D {\bf 85}, 075022 (2012)
  [arXiv:1201.0315 [hep-ph]].
  
\bibitem{Batell:2012ca} 
  B.~Batell, S.~Gori and L.~-T.~Wang,
  arXiv:1209.6382 [hep-ph].
  

\bibitem{Kannike:2011ng} 
  K.~Kannike, M.~Raidal, D.~M.~Straub and A.~Strumia,
  JHEP {\bf 1202}, 106 (2012)
  [arXiv:1111.2551 [hep-ph]].


\bibitem{Dermisek:2013gta} 
  R.~Dermisek and A.~Raval,
  Phys.\ Rev.\ D {\bf 88}, 013017 (2013)
  [arXiv:1305.3522 [hep-ph]].


  
  
  
\bibitem{Dermisek:2014qca} 
  R.~Dermisek, J.~P.~Hall, E.~Lunghi and S.~Shin,
  JHEP {\bf 1412}, 013 (2014)
  [arXiv:1408.3123 [hep-ph]].
  
  
  
  \bibitem{Pierce:1997}
  D.~M.~Pierce, J.~A.~Bagger, K.~T.~Matchev, R.~Zhang,
  Nucl.\ Phys.\ B {\bf 491}, 1 (1997)
  

  

  
  
  
\bibitem{Martin:1993}
S.~P.~Martin and M.~T.~Vaughn,
Phys.\ Rev.\ D {\bf 50}, 2282 (1994)
[hep-ph/9311340]
  

\bibitem{Machacek:1983tz} 
  M.~E.~Machacek and M.~T.~Vaughn,
  Nucl.\ Phys.\ B {\bf 222}, 83 (1983);
 

  
  \bibitem{Jones:1975}
  D.R.T.~Jones
  Nucl.\ Phys. \ B {\bf 87} 127 (1975)


\bibitem{Machacek:1983fi} 
  M.~E.~Machacek and M.~T.~Vaughn,
  Nucl.\ Phys.\ B {\bf 236}, 221 (1984);


  
\bibitem{Castano:1993ri} 
  D.~J.~Castano, E.~J.~Piard and P.~Ramond,
  Phys.\ Rev.\ D {\bf 49}, 4882 (1994)
  [hep-ph/9308335].

  
  
\bibitem{Martin:1993yx} 
  S.~P.~Martin and M.~T.~Vaughn,
  Phys.\ Lett.\ B {\bf 318}, 331 (1993)
  [hep-ph/9308222].

 

\end{thebibliography}
\end{document}